\RequirePackage[l2tabu, orthodox]{nag}
\documentclass{article}
\usepackage[english]{babel}
\usepackage[top=2cm,bottom=2cm,left=2cm,right=2cm]{geometry}
\usepackage{amsmath,amsthm,amssymb,amsfonts}
\usepackage{graphicx} 
\usepackage{syntonly}
\usepackage{subfiles}
\usepackage{caption}
\usepackage{subcaption}
\usepackage{booktabs}
\usepackage{listings}
\usepackage{xcolor}
\usepackage{tabularray}
\usepackage{multirow}
\UseTblrLibrary{siunitx}
\usepackage{algorithmic}
\usepackage{algorithm}
\usepackage{array}
\usepackage{float}
\usepackage{hyperref} 
\usepackage{cleveref}
\usepackage[sort]{natbib}
\setcitestyle{numbers,square,sort}
\usepackage{authblk}

\newtheorem{definition}{Definition}
\newcommand{\refeq}[1]{Eq.(\ref{#1})}
\newcommand{\reffig}[1]{Figure \ref{#1}}
\newcommand{\reftab}[1]{Table \ref{#1}}

\title{\textbf{Systemic Risk Management via Maximum Independent Set in Extremal Dependence Networks}}

\author[a]{Qian Hui \thanks{\href{mailto:qhui24@m.fudan.edu.cn}{qhui24@m.fudan.edu.cn}.}}
\author[a,b]{Tiandong Wang \thanks{Corresponding author, \href{mailto:td_wang@fudan.edu.cn}{td\underline{ }wang@fudan.edu.cn}.}}

\affil[a]{Shanghai Center for Mathematical Sciences, Fudan University}
\affil[b]{Shanghai Academy of Artificial Intelligence for Science}

\date{}
\begin{document}
\maketitle

\begin{abstract}
The failure of key financial institutions may accelerate risk contagion due to their interconnections within the system. In this paper, we propose a robust portfolio strategy to mitigate systemic risks during extreme events. We use the stock returns of key financial institutions as an indicator of their performance, apply extreme value theory to assess the extremal dependence among stocks of financial institutions, and construct a network model based on a threshold approach that captures extremal dependence. Our analysis reveals different dependence structures in the Chinese and U.S. financial systems. By applying the maximum independent set (MIS) from graph theory, we identify a subset of institutions with minimal extremal dependence, facilitating the construction of diversified portfolios resilient to risk contagion. We also compare the performance of our proposed portfolios with that of the market portfolios in the two economies.

\end{abstract}

\section{Introduction}
Financial systems are generally complex and interconnected, making them vulnerable to sudden and extreme fluctuations that may destabilize entire economies. Market crashes and systemic crises often originate from the failure of key financial institutions, such as banks and insurance companies, whose interconnection accelerates risk contagion. 
In the context of insurance, systemic risk is particularly relevant because insurers can contribute to systemic risk through common exposures between companies, which can endanger financial stability in the event of an adverse shock.
Therefore, understanding systemic risk and identifying financial institutions whose distress may spread through the system are the primary concerns of regulators \cite{acharya2010manufacturing, allen2012asset}.

Recent research on systemic risk has increasingly utilized network theory alongside advanced statistical and computational methods to model how shocks spread within financial systems. For instance, \cite{DebtRank} introduces the DebtRank metric to assess the systemic impact of financial institutions, while \cite{AcemogluEtAl:2015} examines how the network of interbank liabilities can amplify risk contagion. Other approaches, such as stress testing and extreme value theory, have been used to quantify tail risks, e.g. the well-known CoVaR framework proposed in \cite{CoVaR} and the investigation in \cite{AcharyaEtAl:2016} on the contribution of financial institutions to systemic risk. Despite these significant advances, how to properly integrate the intertwining financial networks with metrics that can capture the simultaneous occurrence of extreme events still remains under-explored. Successful efforts in this direction can provide a comprehensive understanding of risk propagation in heavy-tailed financial environments and offer valuable insights for developing more robust risk management strategies.

A key challenge in modeling financial systems is accurately capturing dependence structures between institutions, especially under extreme market conditions. Traditional correlation-based methods often break down when dealing with heavy-tailed distributions, since the second moment may not even exist. The limitations of these methods have been well documented in existing work such as \cite{poon2004extreme, kelly2014tail}, leading to the adoption of alternative approaches to measure extremal dependence. Two widely used methods are \emph{extremal dependence measure} (EDM) \cite{resnick2004extremal} and \emph{extremograms} \cite{davis2009extremogram}. EDM quantifies the probability that extremal losses occur simultaneously across institutions, while extremograms track how extreme events propagate over time. A detailed comparison of these methods can be found in \cite{EDM}.

In this study, we analyze the systemic structure of financial systems by quantifying pairwise extremal dependence and representing these relationships as a network. Using the EDM, we construct a network where edges indicate strong extremal dependence between institutions, while the absence of an edge implies little systemic connectivity. To design portfolios resilient to systemic crises, we apply the \emph{maximum independent set} (MIS) from graph theory to our optimization process. The MIS identifies a subset of institutions with minimal extremal dependence, allowing diversification that may reduce exposure to systemic shocks. Although previous work \cite{boginski2005statistical, spelta2022chaos} has demonstrated the value of MIS in strengthening financial networks, the role of extremal dependence in managing risk exposures remains unexplained.

By carefully incorporating the MIS into portfolio construction, our approach in general provides a robust, data-driven method to understand and manage systemic risks. This strategy ensures that portfolios are not only reasonably diversified but also strategically insulated from risk contagion. Our findings provide valuable insights for investors and regulators seeking to enhance financial stability and resilience in an increasingly interconnected global market.

The rest of this paper is organized as follows. Section \ref{sec2} collects key technical concepts including multivariate regular variation and the definition of EDM. Section \ref{sec3} discusses the construction details of extremal dependence networks, analyzes the network structures for both Chinese and U.S. financial systems, and proposes the MIS-based portfolio strategy. Then in Section \ref{sec4}, we compare the performance of the constructed portfolios in both Chinese and U.S. cases, demonstrating the effectiveness of the proposed strategy. Concluding remarks are given in Section \ref{sec5}.

\subsection{Data example}
We now explain the datasets and analytical framework examined throughout the remaining paper.
We use the R package \verb6quantmod6 to retrieve data from Yahoo Finance on 48 stocks from Chinese A-shares and 37 stocks from U.S. S\&P 500 in 2023, all of which are banks and insurance companies. The chosen time interval spans from January 1, 2023 to December 31, 2023, with a total of 242 trading days.
We compute the log-return of stock $i$ on day $t$ as
\[
	r_i(t):=\log P_i(t) -\log P_{i}(t-1),
\]
where $P_{i}(t)$ represents the adjusted closing price of stock $i$ on day $t$. Then we calculate the extremal dependence measure (EDM) between stock $p$ and stock $q$, $\text{EDM}(p,q)$, by substituting the returns into \refeq{eq2.5}. 

Combined with the data above, we outline an algorithm in the following to construct portfolios based on their extremal dependence structures.
\begin{algorithm}[H]
    \caption{Portfolio construction using EDM.}
    \renewcommand{\algorithmicrequire}{\textbf{Input:}}
    \renewcommand{\algorithmicensure}{\textbf{Output:}}
    \begin{algorithmic}
        \REQUIRE Adjusted price of each stock $P_i(t)$, $i=1,\ldots,n$, at time $t$.
        
        Step 1: Compute the log return $r_i(t)$ for stock $i$, and then calculate the pairwise EDM based on \refeq{eq2.5};
        
        Step 2: Denote each stock as a vertex, and use a threshold-based approach to construct networks;
        
        Step 3: Solve for the maximum independent set of each network combining with the vertex centrality;

        Step 4: Use risk measurement indicators such as ES for each maximum independent set, and construct a portfolio optimization model by minimizing the overall risk.
        
        \ENSURE Optimal portfolio with minimum expected shortfalls.
    \end{algorithmic}
\end{algorithm}

\section{Extremal dependence measure}\label{sec2}
The extremal dependence measure (EDM) (cf. \cite{EDM}) quantifies the tendency for large values to occur simultaneously between two components, and we further use it as our main tool to construct the network structure between stock returns.

We start by introducing the definition of regular variation.
In one dimension, a measurable function \(f\) is regularly varying with index $\alpha$, \(\alpha \in \mathbb{R} \)  if $f: \, \mathbb{R}_{+} \mapsto \mathbb{R}_{+}$ satisfies
\begin{equation}
	\lim_{t\rightarrow \infty} \frac{f(tx)}{f(t)} = x^{\alpha}, \,\, \text{for} \,\, x>0,
\end{equation}
denoted as $f \in RV_{\alpha}$. To formalize our analysis, we provide some useful definitions related to multivariate regular variation (MRV) of measures, and it is a natural extension of the one-dimensional regular variation. 

Suppose that $\mathbb{C}_0 \subset \mathbb{C} \subset \mathbb{R}_{+}^2$ are two closed cones, and we provide the definition of $\mathbb{M}$-convergence in Definition \ref{def1} (cf. \cite{basrak2019note, das2013living, hult2006regular, kulik2020heavy, Lindskog2013RegularlyVM}) on $\mathbb{C} \setminus \mathbb{C}_0$, which lays the theoretical foundation of regularly varying measures (cf. Definition \ref{def2}).

\begin{definition}\label{def1}
Let $\mathbb{M}(\mathbb{C} \setminus \mathbb{C}_0)$ be the set of Borel measures on $\mathbb{C} \setminus \mathbb{C}_0$ which are finite on sets bounded away from $\mathbb{C}_0$, and $\mathbb{C} \setminus \mathbb{C}_0$ be the set of continuous, bounded, non-negative functions on $\mathbb{C} \setminus \mathbb{C}_0$ whose supports are bounded away from $\mathbb{C}_0$. Then for $\mu_n$, $\mu \in \mathbb{M}(\mathbb{C} \setminus \mathbb{C}_0)$, we say $\mu _n \rightarrow \mu $ in $\mathbb{M}(\mathbb{C} \setminus \mathbb{C}_0)$, if $\int f\text{d}\mu_n \rightarrow \int f\text{d}\mu$ for all $f \in \mathcal{C} (\mathbb{C} \setminus \mathbb{C}_0)$.
\end{definition}

\begin{definition}\label{def2}
The distribution of a random vector $\textbf{Z}=[Z_1,Z_2]^T$ on $\mathbb{R}_{+}^2$, i.e. $\mathbb{P}(\textbf{Z} \in \cdot)$, is (standard) regularly varying on $\mathbb{C} \setminus \mathbb{C}_0$ with index $c>0$ (written as $\mathbb{P}(\textbf{Z} \in \cdot) \in \text{MRV}(c,b(t),\nu,\mathbb{C} \setminus \mathbb{C}_0)$) if there exists some scaling function $b(t) \in RV_{1/c}$ and a limit measure $\nu(\cdot) \in \mathbb{M}(\mathbb{C} \setminus \mathbb{C}_0)$ such that as $t\rightarrow \infty$,
\begin{equation}\label{eq2.1}
    t\mathbb{P}\left( \frac{\textbf{Z}}{b(t)} \in \cdot\right)  \rightarrow \nu(\cdot), \,\, \text{in} \,\, \mathbb{M}(\mathbb{C} \setminus \mathbb{C}_0).
\end{equation}
\end{definition}
In \refeq{eq2.1}, all elements are normalized by the same function column $b(t)$, which implies that all marginal distributions are tail-equivalent with index $-\alpha$ \cite{EDMbook}. When analyzing the asymptotic dependence between components of a bivariate random vector $\textbf{Z}$ satisfying \refeq{eq2.1}, it is often informative to make a polar coordinate transform and consider the transformed points located on the $L_2$ unit sphere
\begin{equation}
    (x,y) \mapsto \left( \frac{x}{\sqrt{x^2+y^2}}, \frac{y}{\sqrt{x^2+y^2}} \right) ,
\end{equation}
after thresholding the data according to the $L_2$ norm. In $\mathbb{R}_{+}^2$, the convenient version of the $L_2$-polar coordinate transformation is $T:\, \textbf{Z} \mapsto (\lVert \textbf{Z} \lVert , \textbf{Z} / \lVert \textbf{Z} \lVert ) = (R, \Theta)$, we provide the following equivalent definition in polar coordinates. 

\begin{definition}(cf. \cite[Theorem 6.1]{EDMbook}) 
A 2-dimensional random vector $\textbf{Z}=[Z_1,Z_2]^T$ is (standard) regularly varying if and only if there exists a function sequence $b(t) \rightarrow \infty$ and a spectral measure $\Gamma$ on $\aleph _{+}^2 = \{\textbf{x} \in \mathbb{R}_{+}^2 \setminus \{0\} :\, \lVert \textbf{x} \lVert = 1\}$, and there exists a constant $c = \nu \{\textbf{x}: \lVert \textbf{x} \lVert > 1\} > 0$ such that
	\begin{equation}\label{eq2.2}
		t\mathbb{P}\left( \left( \frac{R}{b(t)}, \Theta \right)  \in \cdot\right)  \rightarrow c\nu_{\alpha} \times \Gamma, \,\, \text{in} \,\, \mathbb{M}_{+}(\left( 0, \infty \right] \times \aleph_{+}^2),
	\end{equation}
where $\nu_{\alpha} \left( x, \infty \right] = x^{-\alpha}$, $x>0$.
\end{definition}


Now we focus on the extremal dependence measure. Given a regularly varying bivariate random vector $\textbf{Z} = [Z_1,Z_2]^T$, the EDM is defined as (cf. \cite{EDM}, Eq.(8))
\begin{equation}\label{eq2.3}
	\text{EDM}(Z_1,Z_2) = \int _{\aleph^2_{+}} a_1a_2 \Gamma (d\textbf{a}).
\end{equation}
Notice that the value of EDM is 0 if and only if the coordinates of $\textbf{Z}$ are asymptotically independent, i.e., the spectral measure $\Gamma$ concentrates on $\{ (1,0) / \lVert (1,0) \lVert , (0,1) / \lVert (0,1) \lVert \}$, or equivalently, the limit measure $\nu$ concentrates on the axes. In addition, if the norm is symmetric, then EDM reaches its maximum value if and only if the support of $\Gamma$ is $\{\textbf{a}:\, a_1=a_2 \}$, or equivalently, $\nu$ concentrates on the line $\{t(1,1),t>0\}$. 

In \cite{EDM}, the authors highlight that $\text{EDM}$ can be interpreted as the limit of the cross moment between normalized $Z_1$ and $Z_2$ when $R = \lVert \textbf{Z} \lVert $ is large, i.e.
\begin{equation}\label{eq2.4}
	\text{EDM}(Z_1,Z_2) = \mathop{\lim}\limits_{x \rightarrow \infty } \mathbb{E} \left[ \frac{Z_1}{R} \frac{Z_2}{R} \middle\vert R>x \right]. 
\end{equation}
Based on this relationship, they proposed an estimator for $\text{EDM}(Z_1,Z_2)$, which is defined as
\begin{equation}\label{eq2.5}
	\widehat{\text{EDM}} (Z_1,Z_2) = \frac{1}{N_n} \mathop{\sum}\limits_{i=1}^n \frac{Z_{i1}}{R_i} \frac{Z_{i2}}{R_i} {\textbf{1}_{[R_i \geq x ]}},
\end{equation}
where $\textbf{Z}_i = [Z_{i1},Z_{i2}]^T$ $(1\leq i \leq n)$ is iid random vector, $R_i = \lVert \textbf{Z}_i \lVert $, and $N_n = \mathop{\sum}\limits_{i=1}^n {\textbf{1}_{[R_i \geq x ]}}$. Note that \refeq{eq2.5} suggests the value range of $\text{EDM}$ is $[-0.5, 0.5]$. 
In the next section, we will construct extremal dependence networks among stock returns using EDM as the main character.

\section{Stock network model based on extremal dependence}\label{sec3}
In this section, we use EDM to construct a network that describes the pairwise extremal dependence structure of the stock returns. By specifying such a network structure, we later develop stock selection strategies in Section~\ref{sec4}.
A complex network consists of a set of vertices $V$ and a set of edges $E$, denoted as $G=(V,E)$. An undirected edge connecting the vertices $i$ and $j$ is represented as $\{i, j\}$. 
We start by summarizing important network characteristics and then discuss how to construct a network using EDMs.

\subsection{The statistical properties of the network}\label{subsec:network}
Complex networks analyze the properties of vertices and edges from a statistical perspective and can describe the characteristics of a network from various aspects. Here we focus on the following six properties and use them in Section~\ref{sec3.2} to compare network characteristics at different thresholds. This analysis will help identify the most suitable threshold for network construction.

\subsubsection{Average degree and degree distribution}\label{sec:degree}
The vertex degree refers to the number of edges connected to a given vertex. The average degree is the mean of the degrees of all vertices in the network and is generally used to measure the overall level of connectivity among vertices. A higher average degree indicates that the edges are more closely connected within the network, suggesting a higher level of interconnection. Degree distribution describes the distribution of degrees among the vertices in the network. If the degree distribution follows a power law, it means that a few vertices have high degrees, while most vertices have lower degrees. Scale-free networks exhibit this property. The degree distribution of a scale-free network is typically represented in a power-law form, as follows (cf. \cite[Chap.1 p.3]{van2024random}) 
\begin{equation}
	\mathbb{P}(n) \propto n^{-\alpha},
\end{equation}
where $\mathbb{P}(n)$ represents the probability density of the \( n \)th vertex, with \( \alpha \) as the estimated parameter.

\subsubsection{Average path length} 
The average path length refers to the mean distance between any two vertices in a network, where distance is typically defined as the minimum number of edges needed to connect the two vertices. A shorter average path length indicates that vertices in the network can influence each other more readily, and information can spread more efficiently across the network. Average path length is a crucial metric for measuring the overall connectivity and efficiency of a network. The calculation formula is as follows (cf. \cite[Chap.1 p.4]{van2024random})
\begin{equation}
	L = \frac{2}{N(N-1)} \sum _{i\geq j}d_{ij},
\end{equation}
where \(N\) denotes the number of vertices, and \(d_{ij}\) represents the number of edges between vertices \(i\) and \(j\).

\subsubsection{Clustering coefficient}
The clustering coefficient measures the degree of clustering or cohesion among vertices in a network. It is defined as the probability that any two neighbors of a given vertex are connected. This is calculated as the ratio of the actual number of connections between neighboring vertices to the maximum possible number of connections between them. The formula to calculate the clustering coefficient is as follows (cf. \cite[Chap.1 p.17]{van2024random})
\begin{equation}
	C_{i} = \frac{2L_i}{k_i (k_i-1)},
\end{equation}
where \(L_i\) represents the actual number of connections between the neighboring vertices of vertex \(i\), and \(k_i\) denotes the number of neighboring vertices for vertex \(i\).

\subsubsection{Network diameter}
The network diameter refers to the maximum distance between any two vertices in a network, where distance is defined as the number of edges that must be traversed to connect the two vertices. Network diameter is defined as (cf. \cite[Chap.1 p.4]{van2024random})
\begin{equation}
	D = \max (d_{ij}).
\end{equation}

\subsubsection{Graph density}
The graph density is the ratio of actual connections in a network to the total possible connections. It reflects the level of connectivity in the network. The calculation formula is as follows (cf. \cite[Chap.1 p.42]{van2024random})
\begin{equation}
	\rho = \frac{2M}{N(N-1)},
\end{equation}
where \(M\) represents the actual number of edges in the network.

\subsection{Network construction and important characteristics}\label{sec3.2}
To construct the network of stocks, we denote each stock as a vertex and use a threshold-based approach to construct networks. Here networks constructed under different thresholds have the same number of vertices but differ in the number of edges. In particular, we define the set of edges $E$ as
\begin{equation}
    E = 
    \left\{ \begin{array}{l} 
    e_{ij}=1,\ \ i\ne j \ \ \text{and} \ \ \text{EDM}(i,j)\ge \theta  \\  
    e_{ij}=0,\ \ \text{otherwise} .
    \end{array}\right.
\end{equation}
In other words, the higher the chosen threshold $\theta$ , the sparser the network.

\subsubsection{Choice of threshold}
First of all, we need to select an appropriate threshold $\theta$. Too low a chosen threshold may give a dense network with numerous weak connections, leading to a lack of clear structure. On the other hand, an excessively high threshold isolates many vertices, potentially omitting critical dependencies. The challenge lies in striking a balance between preserving essential dependence structures and avoiding unnecessary complexity.

The resulting Chinese network consists of 48 vertices with EDM values ranging from 0.0217 to 0.5, and the U.S. network consists of 37 vertices with EDM values between -0.0014 and 0.5.
Following the principles above, we examine four threshold values: $\theta = 0.18$, $0.20$, $0.22$, and $0.24$. For each $\theta$, we construct a corresponding network and analyze key characteristics summarized in Section~\ref{subsec:network}. 

\begin{figure}[H]
	\centering
	\subcaptionbox{Threshold=0.18}{
		\begin{minipage}[h]{.23\linewidth}
			\centering
			\includegraphics[scale=0.18]{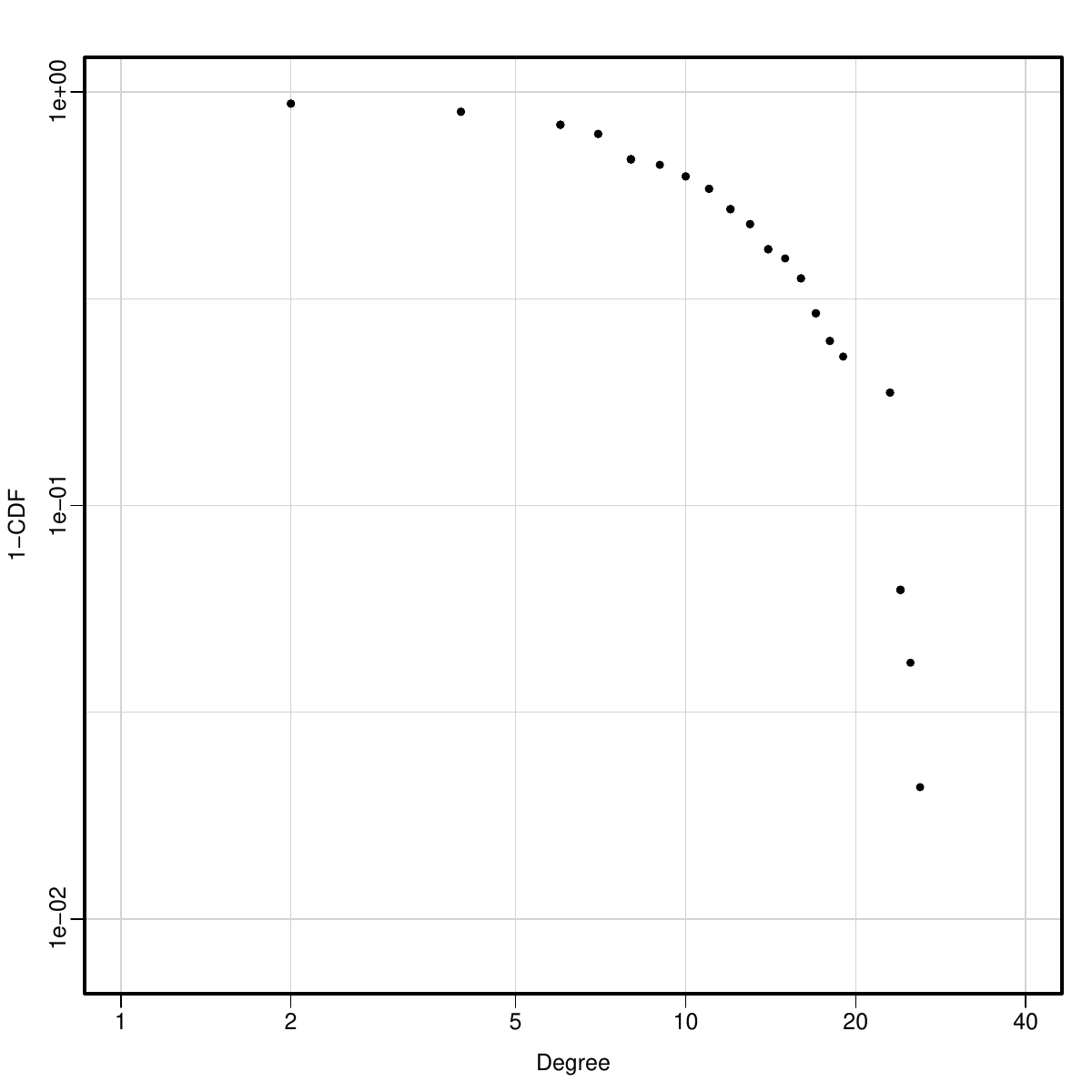}
		\end{minipage}
	}
	\subcaptionbox{Threshold=0.20}{
		\begin{minipage}[h]{.23\linewidth}
			\centering
			\includegraphics[scale=0.18]{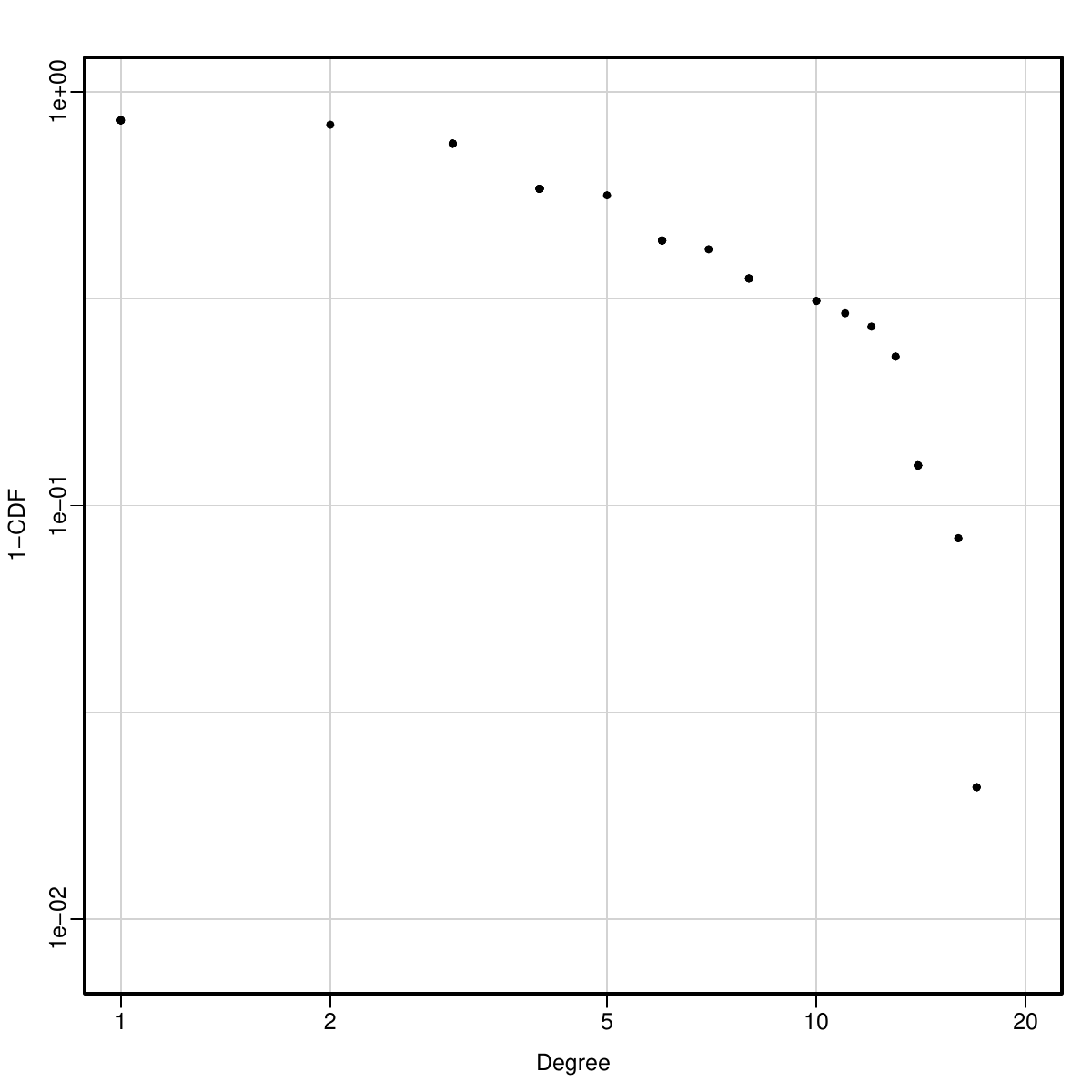}
		\end{minipage}
	}
	\subcaptionbox{Threshold=0.22}{
		\begin{minipage}[h]{.23\linewidth}
			\centering
			\includegraphics[scale=0.18]{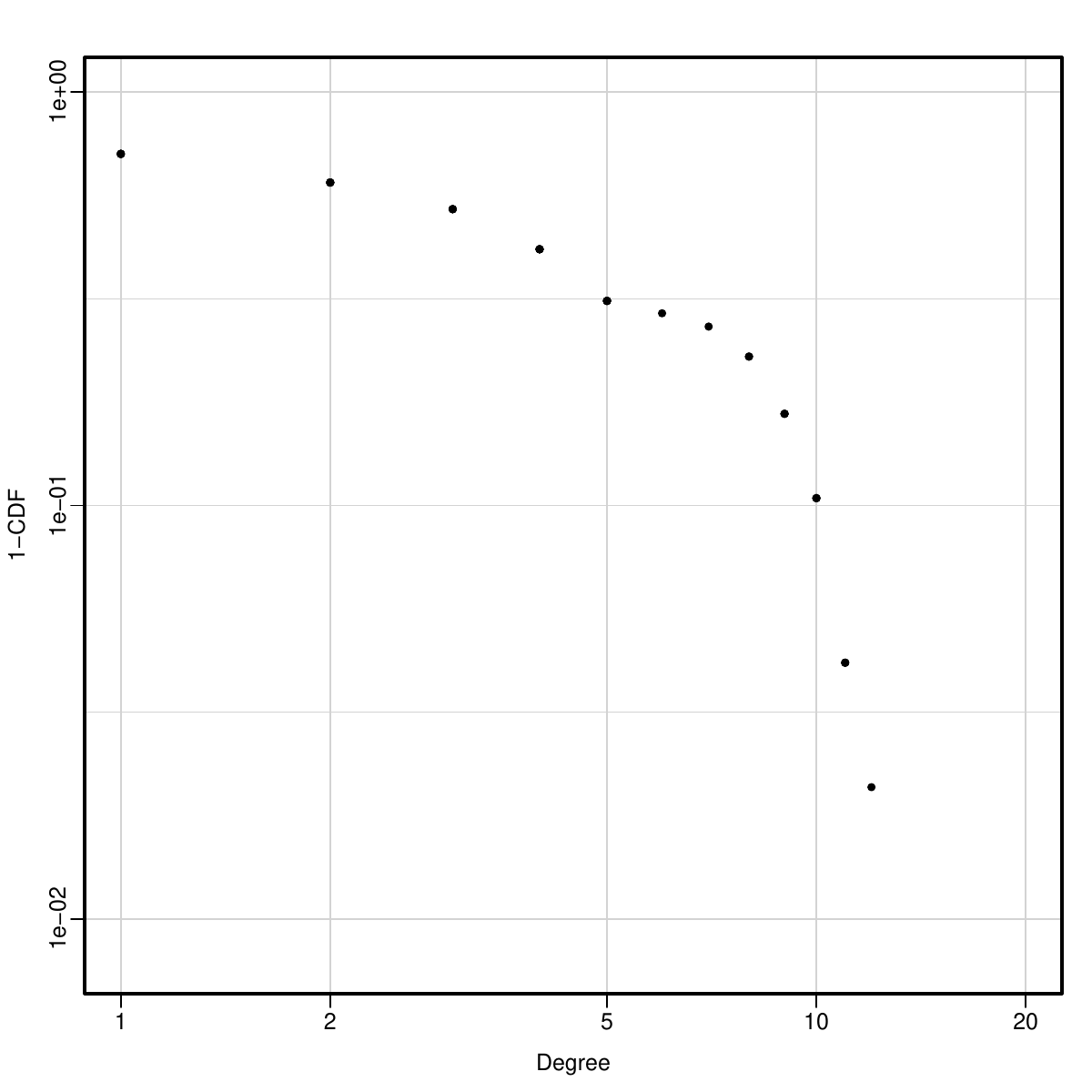}
		\end{minipage}
	}
	\subcaptionbox{Threshold=0.24}{
		\begin{minipage}[h]{.23\linewidth}
			\centering
			\includegraphics[scale=0.18]{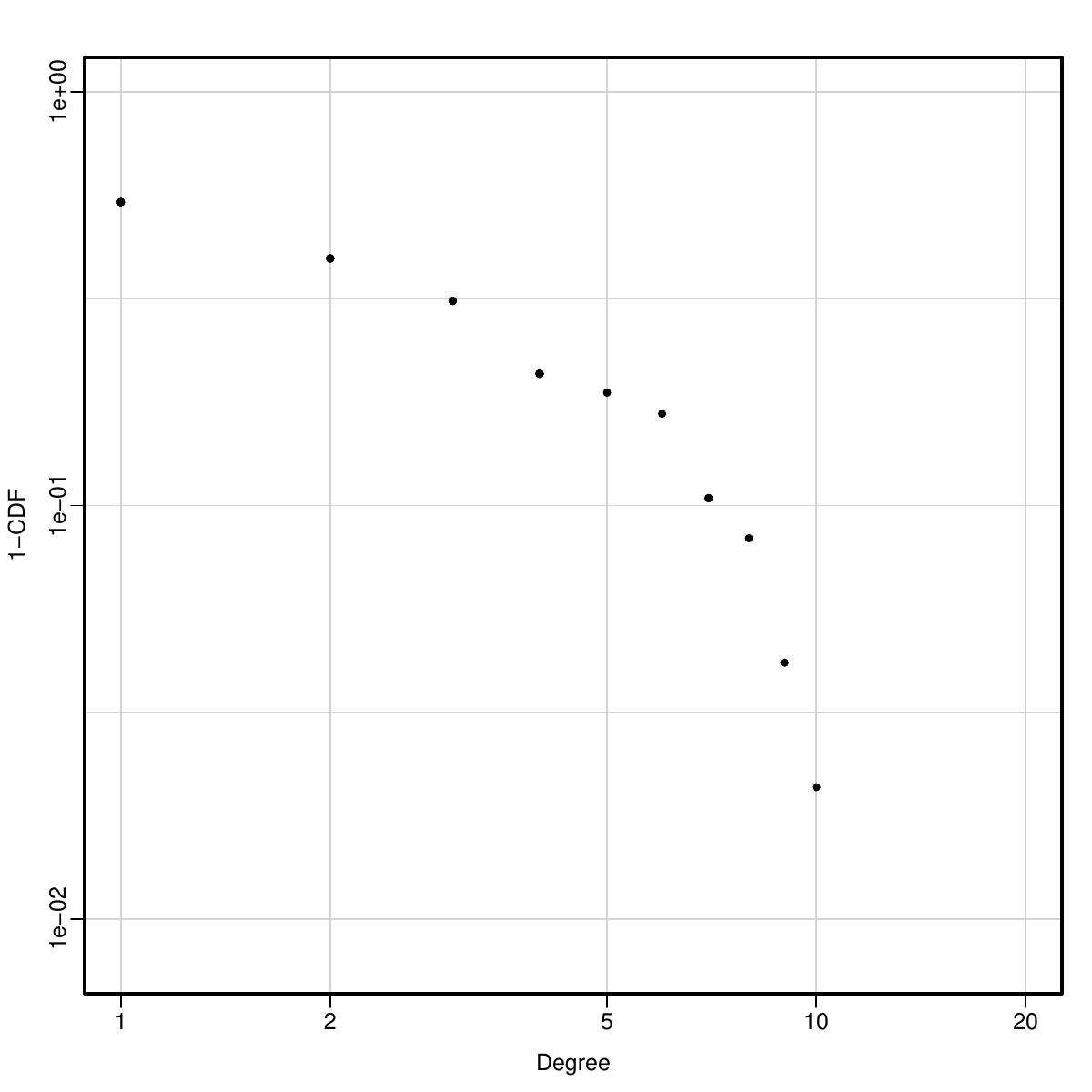}
		\end{minipage}
	}
	\caption{Log-log plots of the complementary cumulative distribution function (1-CDF) for the degrees at different thresholds in the Chinese A-shares market.}
	\label{fig:thred_zh}
\end{figure}
\begin{figure}[H]
	\centering
	\subcaptionbox{Threshold=0.18}{
		\begin{minipage}[h]{.23\linewidth}
			\centering
			\includegraphics[scale=0.18]{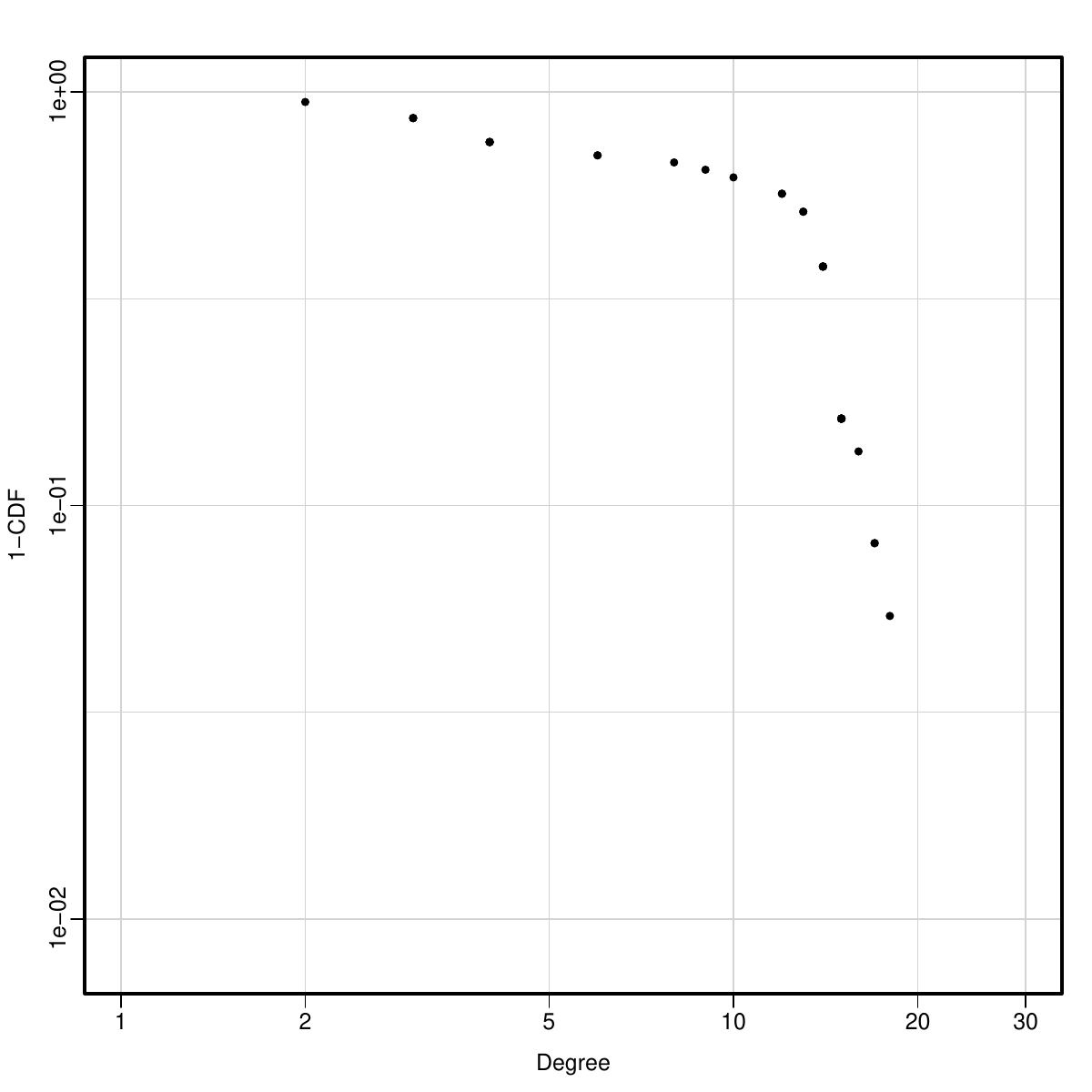}
		\end{minipage}
	}
	\subcaptionbox{Threshold=0.20}{
		\begin{minipage}[h]{.23\linewidth}
			\centering
			\includegraphics[scale=0.18]{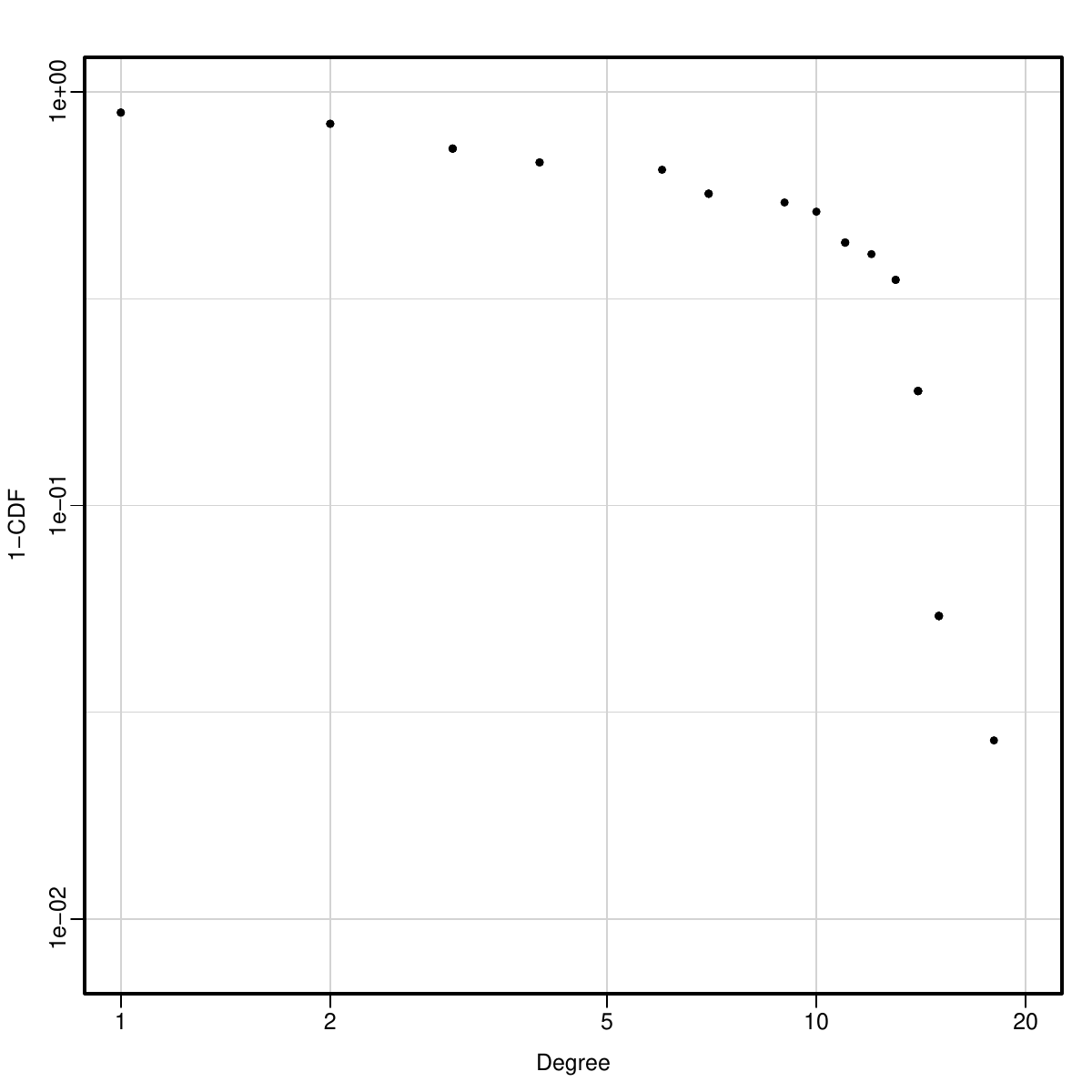}
		\end{minipage}
	}
	\subcaptionbox{Threshold=0.22}{
		\begin{minipage}[h]{.23\linewidth}
			\centering
			\includegraphics[scale=0.18]{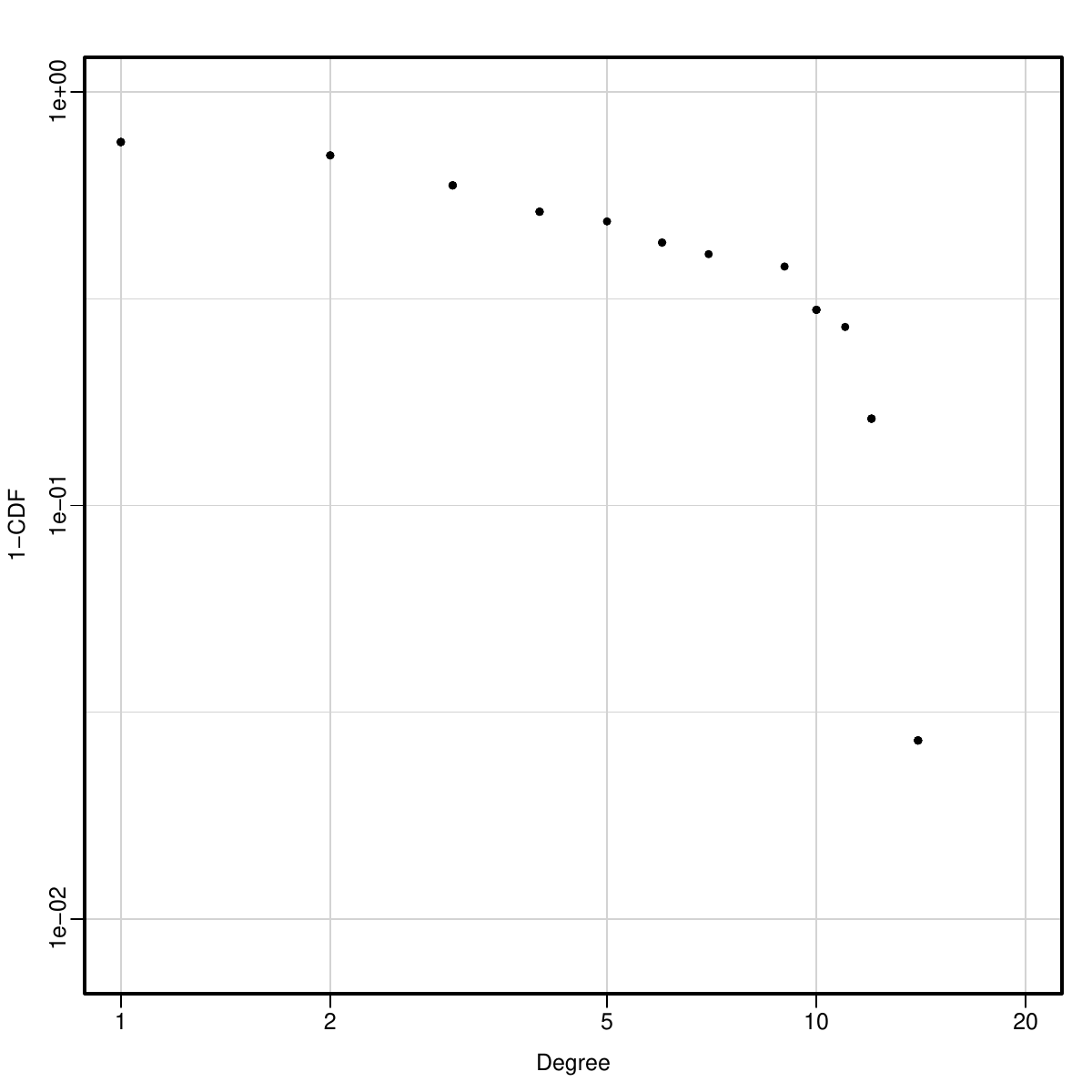}
		\end{minipage}
	}
	\subcaptionbox{Threshold=0.24}{
		\begin{minipage}[h]{.23\linewidth}
			\centering
			\includegraphics[scale=0.18]{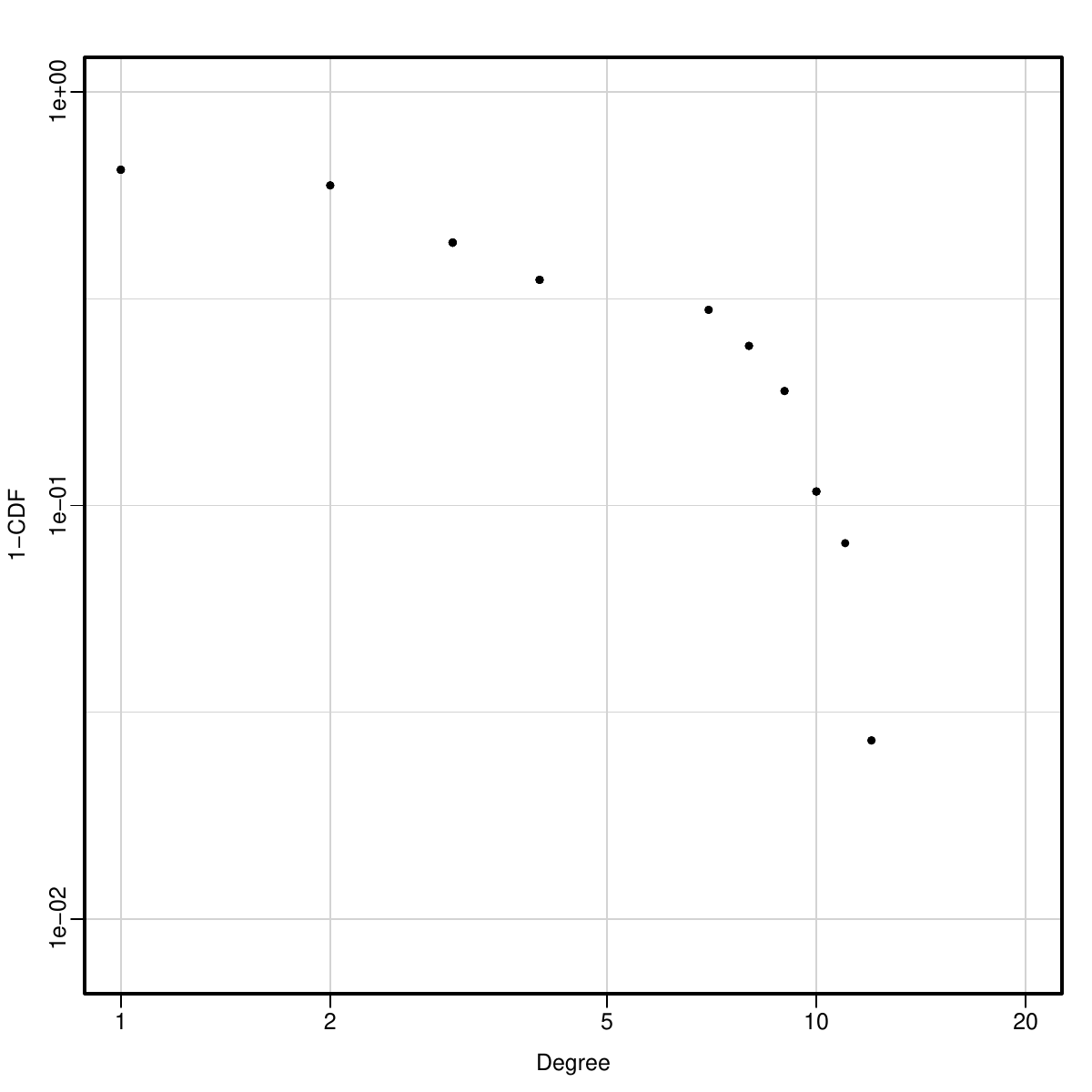}
		\end{minipage}
	}
	\caption{Log-log plots of the complementary cumulative distribution function (1-CDF) for the degrees at different thresholds in the U.S. S\&P 500 market.}
	\label{fig:thred_en}
\end{figure}

We start by examining the empirical degree distributions. In the Chinese A-shares market (Figure~\ref{fig:thred_zh}), networks with $\theta = 0.18$ and $0.20$ (panels (a) and (b)) show rapid decay, with few high-degree nodes and no scale-free behavior. When $\theta = 0.24$, excessive edge loss fragments the network, whereas for $\theta = 0.22$, \reffig{fig:thred_zh}(c) reveals a power-law decay in the degree tail.
In the U.S. S\&P 500 market (\reffig{fig:thred_en}), the trends are similar: cases where $\theta = 0.18$ and $0.20$ yield steep decays, while $\theta = 0.22$ produces a power-law tail, despite minor differences in decay rate and degree range. 

We also inspect other important network metrics for different values of $\theta$.
Table~\ref{tab:thred_zh} presents the results for Chinese A-shares (bank and insurance stocks), while those for the U.S. S\&P 500 (bank and insurance stocks) are summarized in Table~\ref{tab:thred_en}.

\begin{table}[H]
	\centering
	\caption{Network parameters under different threshold values in the Chinese A-shares market (bank and insurance stocks).}
        \tabcolsep=0.45cm
	\begin{tabular}{ccccccc}
		\toprule
		\multirow{2}{*}{Threshold}& Isolated  & Average  & Network & Graph  & Average clustering  & Average  \\
            & vertex& degree& diameter& density& coefficient&path length\\
        \midrule
		0.18& 1& 13.83333& 0.95376& 0.29433& 0.60473& 0.37659\\
		0.20& 3& 7.58333& 1.29882& 0.16135& 0.59502& 0.53945\\
        0.22& 9& 4.62500& 2.24227& 0.09840& 0.67266& 0.91950\\
        0.24&17& 2.70833& 1.50024& 0.05762& 0.69436& 0.54731\\
		\bottomrule
	\end{tabular}
	\label{tab:thred_zh}
\end{table}
\begin{table}[H]
	\centering
	\caption{Network parameters under different threshold values in the U.S. S\&P 500 market (bank and insurance stocks).}
        \tabcolsep=0.45cm
	\begin{tabular}{ccccccc}
		\toprule
		\multirow{2}{*}{Threshold}& Isolated  & Average  & Network & Graph  & Average clustering  & Average  \\
            & vertex& degree& diameter& density& coefficient&path length\\
        \midrule
        0.18& 1& 11.35135& 0.88581& 0.31532& 0.76416& 0.43120\\
		0.20& 2& 9.18919& 1.12520& 0.25526& 0.78852& 0.43942\\
        0.22& 5& 6.48649& 1.16454& 0.18018& 0.75648& 0.48309\\
        0.24& 9& 4.43243& 1.05684& 0.12312& 0.74507& 0.51554\\
		\bottomrule
	\end{tabular}
	\label{tab:thred_en}
\end{table}

According to Table~\ref{tab:thred_zh}, in the Chinese market, as the threshold increases, the number of isolated vertices in the network rises while the number of edges decreases, leading to lower average degrees. For networks with $\theta = 0.18, 0.20, 0.22$, the network diameter, graph density, and average path length all show a positive association with the increasing threshold. Meanwhile, increases in network diameter and average path length reflect reduced network connectivity, while the decreasing graph density indicates a sparser network. We also observe that the network diameter peaks at $\theta = 0.22$, after which it declines due to a growing number of isolated vertices. Since the diameter of a network quantifies the longest effective paths among connected components, it also indicates that financial shocks may take longer to spread.

Although a similar pattern appears in the U.S. market (cf. Table~\ref{tab:thred_en}), different numerical values and sensitivities to the threshold are observed. Overall, we find that the increase in the threshold quickly leads to a large number of isolated vertices in the Chinese market network, while the U.S. networks consistently exhibit smaller network diameters, higher graph densities, and higher average clustering coefficients. These characteristics suggest that U.S. financial institutions may be more vulnerable to systemic risks, as financial shocks or crises could spread more quickly and widely in a more interconnected system.

Considering all characteristics above, we proceed with $\theta = 0.22$ for both China and the U.S.. Then, we visualize the graphs for both the Chinese A-shares and the U.S. S\&P 500 markets. As shown in \reffig{fig:network_zh}(a) and \reffig{fig:network_en}(a), both networks exhibit clear community structures.
In the sequel, we refer to such graphs as \emph{extremal dependence networks} for stocks, and if no edge has been observed between nodes $i$ and $j$, then it means that the corresponding two stock returns show asymptotic independence. We will further analyze the properties of this network in Sections~\ref{subsec:community} and \ref{subsec:portfolio}.

\subsubsection{Community structure of the extremal dependence network via GN algorithm}\label{subsec:community}
Community structure is a key feature of complex networks, where vertices naturally group together, forming smaller, densely connected communities. Within these communities, internal connections are dense, while connections between communities are sparse, revealing a modular organization. This suggests that vertices within the same community have close relationships, likely due to similar characteristics or roles. The concept of community was first introduced in sociology, and it has since found extensive applications across various disciplines, including physics, biology, electronics, and computer science (cf. \cite{bazzi2016community, fortunato2016community, isogai2014clustering, isogai2017dynamic}).

\begin{figure}[h]
	\centering
	\subcaptionbox{}{
		\begin{minipage}[h]{.6\linewidth}
			\centering
			\includegraphics[scale=0.35]{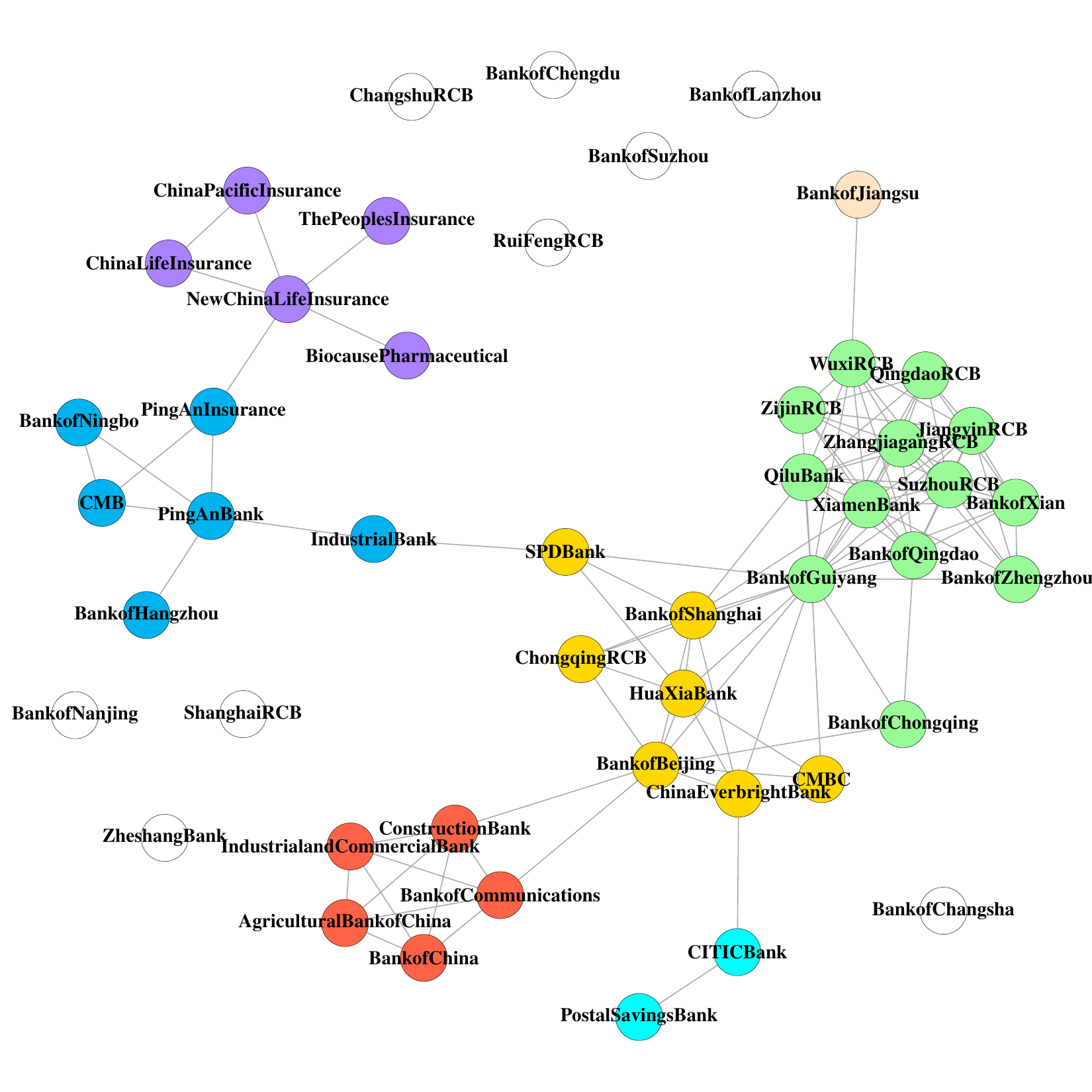}
		\end{minipage}
	}
	\subcaptionbox{}{
		\begin{minipage}[h]{.35\linewidth}
			\centering
			\includegraphics[scale=0.3]{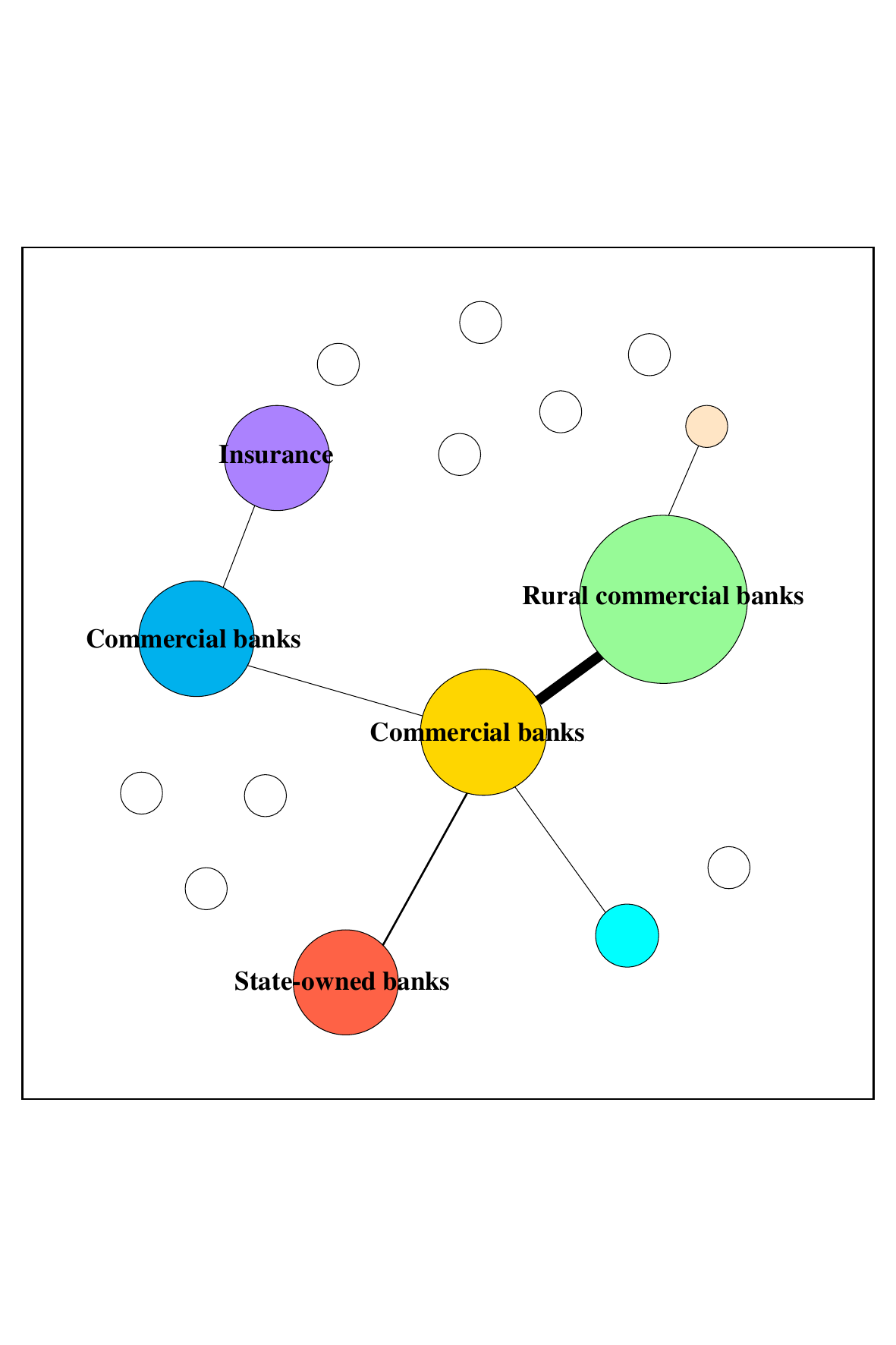}
		\end{minipage}
	}
	\caption{Community structures in extremal dependence networks. (a) The GN algorithm partitions 48 vertices into 16 communities in the Chinese A-shares market, with vertex colors marking community affiliation. (b) A coarse-grained view condenses each community into a single node whose diameter reflects its vertex count, while edges denote collaborations between communities.}
	\label{fig:network_zh}
\end{figure}

\begin{figure}[h]
	\centering
	\subcaptionbox{}{
		\begin{minipage}[h]{.6\linewidth}
			\centering
			\includegraphics[scale=0.35]{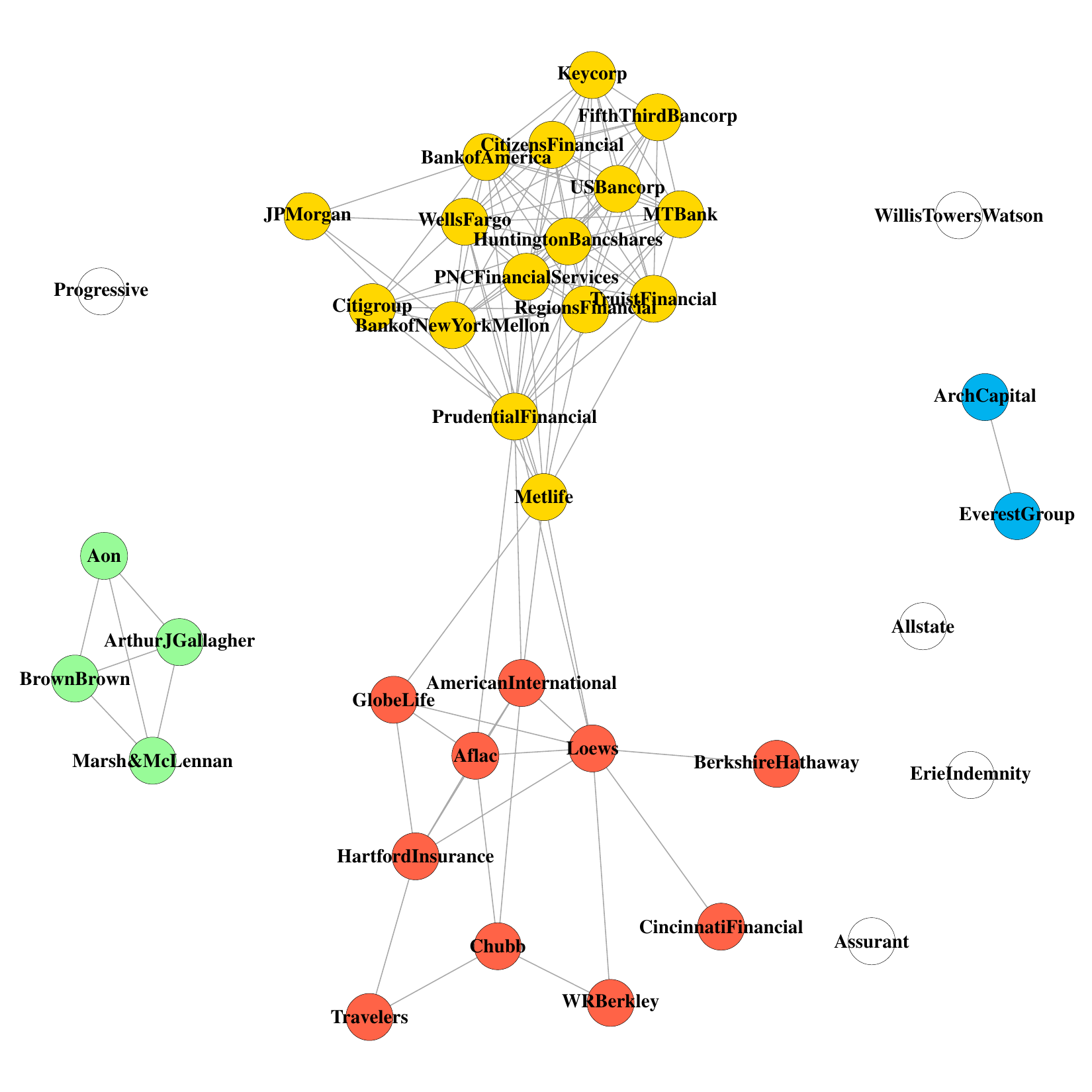}
		\end{minipage}
	}
	\subcaptionbox{}{
		\begin{minipage}[h]{.35\linewidth}
			\centering
			\includegraphics[scale=0.3]{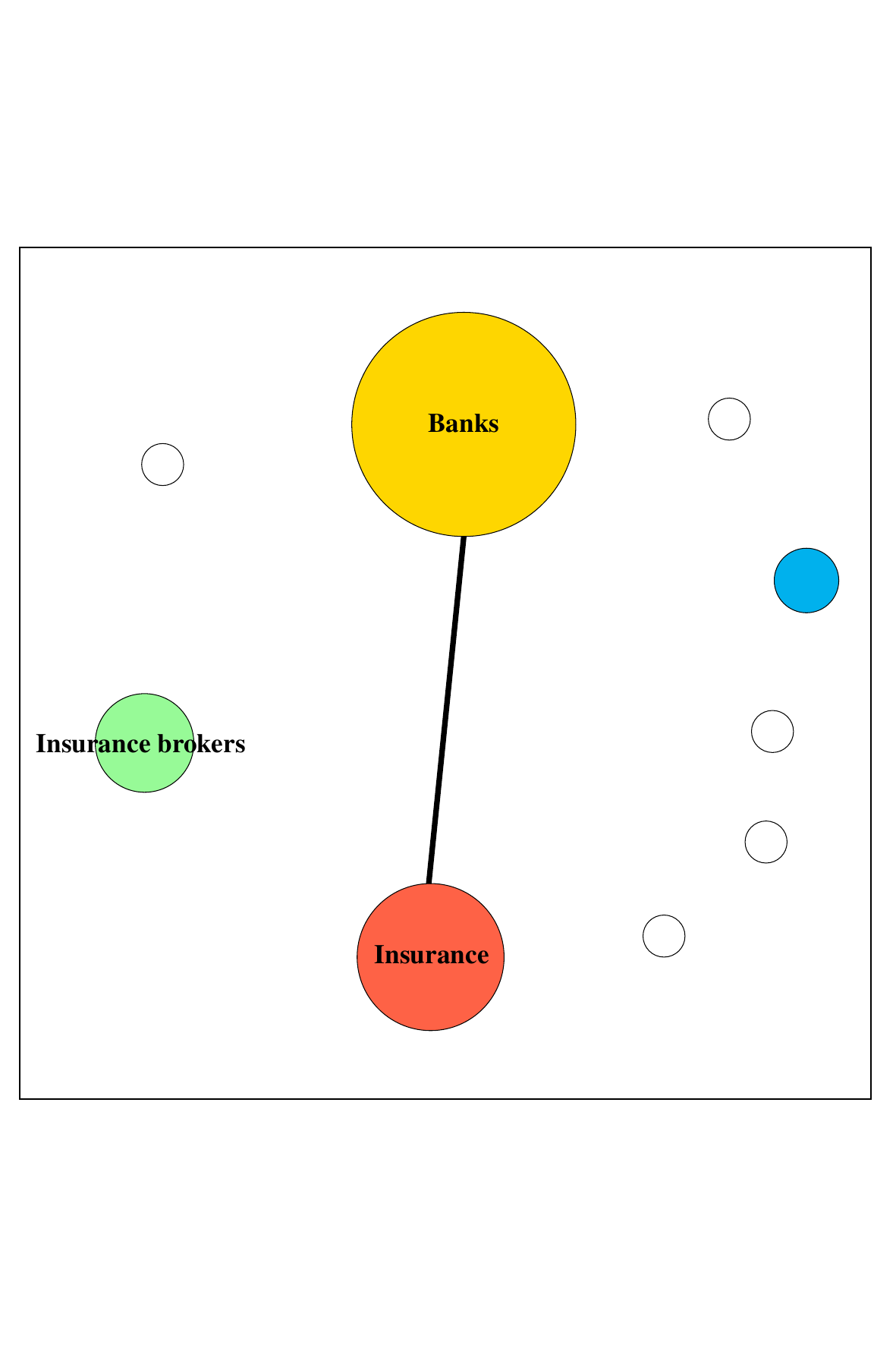}
		\end{minipage}
	}
	\caption{Community structures in extremal dependence networks. (a) The GN algorithm partitions 37 vertices into 9 communities in the U.S. S\&P 500 market, with vertex colors marking community affiliation. (b) A coarse-grained view condenses each community into a single node whose diameter reflects its vertex count, while edges denote inter-community collaborations between communities.}
	\label{fig:network_en}
\end{figure}

Among the numerous algorithms for identifying community structures in networks (cf. \cite{kernighan1970efficient, capocci2005detecting, pothen1990partitioning, clauset2004finding, newman2004fast}), we use the Girvan-Newman (GN) algorithm, which is particularly effective for uncovering hierarchical divisions. Using GN, we partition the Chinese and U.S. extremal dependence networks into 16 and 9 communities, respectively. Note that each isolated vertex is considered an individual community.
The graphs with community divisions, labeled by different colors for both China and the U.S., are shown in \reffig{fig:network_zh}(a) and \reffig{fig:network_en}(a), respectively.

Furthermore, within each community that contains at least two nodes, we aggregate the financial institutions in that community and plot the aggregated graphs in \reffig{fig:network_zh}(b) and \reffig{fig:network_en}(b), where larger vertices indicate a larger community size.
The edges between communities indicate their collaborations, with the thickness of the edges varying in proportion to the number of edges between them. These aggregated panels illustrate the high-level structure of the extremal dependence network, revealing different interactions between communities.

In the Chinese extremal dependence network (\reffig{fig:network_zh}), we first note that the five state-owned banks (namely, Industrial and Commercial Bank of China, Construction Bank, Bank of China, Agricultural Bank of China, and Bank of Communications) form a completely connected community (colored red), which is directly linked to one of the two commercial bank communities (yellow). In particular, this interaction is facilitated through the triangle formed by the Construction Bank, Bank of Communications, and Bank of Beijing. This triadic closure indicates strong extremal dependence among the three banks, positioning them as key transmission vertices if any extreme events occur in the state-owned banks. Such dependence may be due to their shared regional focus in Beijing, the capital of China, and similar exposure to local government projects.

The yellow community forms two major connections: one to the communities of rural commercial banks (green) and another to a different group of commercial banks (blue). Comparing the yellow and blue clusters, we observe that the yellow one is more regionally focused, while the blue one is more nationally and internationally oriented, with diversified business models including wealth management and investment banking. Regional commercial banks show a notably higher density of connections with rural commercial banks, suggesting strong inter-business ties among them. Within the blue cluster, Ping An Bank and Ping An Insurance are connected, exemplifying their affiliation with the Ping An Group. Also, through Ping An Insurance, a purple community emerges, consisting of major insurance companies in China. Most isolated vertices are local commercial banks that exhibit no extremal dependence on other institutions.

Instead of the hierarchical structure seen in the Chinese extremal dependence network, \reffig{fig:network_en} shows a very different pattern in the U.S. case. The aggregated plot (cf. \reffig{fig:network_en}(b)) depicts that banks and insurance companies form two dominant communities, which are primarily linked by two key intermediaries: Prudential Financial and MetLife. This coincides with the designation by the Financial Stability Board \cite{FSB2019}, which identifies Prudential Financial and MetLife as two of the top 10 globally systemically important insurers. 
The bank community (yellow) shows dense connections within itself and includes the ``big four" banks in the U.S. system (i.e., JPMorgan Chase, Bank of America, Citigroup, and Wells Fargo). However, the insurance community (red) is sparser. Additionally, a separate cluster of insurance brokers forms a completely connected subgraph, but it is not linked to any of the other clusters, underscoring the industry's strong internal connectivity.

In conclusion, by examining the community structures of the two extremal dependence networks, we see that the Chinese and U.S. financial network systems reveal key structural differences that may influence their systemic risk profiles. The Chinese network is hierarchical, with tightly interconnected state-owned banks that share regional and government ties. However, the U.S. network is somewhat bipartite, with banking and insurance communities linked by key intermediaries like Prudential Financial and MetLife. 
Overall, the Chinese system may be more vulnerable if extreme shocks take place within its state-owned banks, while the U.S. system may face cross-sectoral contagion from failures of key intermediaries.


\subsubsection{Centrality measure of the extremal dependence network}\label{subseck:btw}

Apart from the community structure, \emph{betweenness centrality} is another important measure for the constructed extremal dependence network since it identifies key vertices controlling the spread of systemic risks. Vertices with high betweenness are potentially vulnerable points, where failure may trigger widespread financial contagion.

In network science, the betweenness centrality of a vertex $v$ is defined as
\begin{equation*}
    B(v) = \sum_{1 \leq i < j \leq n, i\neq v, j\neq v} n^v_{ij}/n_{ij},
\end{equation*}
where $n_{ij}$ is the total number of shortest paths between vertices $i$ and $j$, $n^v_{ij}$ is the number of those shortest paths which pass through vertex $v$, and $n$ is the number of vertices in the network. Also, the normalized betweenness centrality is given by
\begin{equation*}
    B_{N}(v) = \frac{2B(v)}{(n-1)(n-2)}.
\end{equation*}
We now list the top eight stocks with the highest betweenness centralities in \reftab{tab:betweenness_zh} and \reftab{tab:betweenness_en}.
Combining the tables with Figures \ref{fig:network_zh} and \ref{fig:network_en}, we see that vertices with high betweenness are those linking different communities. For regulators, these nodes are of particular concern, as their failure may trigger systemic risk contagion, spreading instability across the network. 

\begin{table}[H]
	\centering
	\caption{The top 8 stocks by vertex betweenness centrality in the Chinese network.}
	\tabcolsep=0.38cm
	\begin{tabular}{ccccc}
		\toprule
		  & Bank of Guiyang& SPD Bank& Industrial Bank& Ping An Bank  \\
		\midrule
        $B$ & 369& 297& 280& 269 \\
		$B_{N}$  & 0.3414& 0.2747& 0.2590& 0.2488 \\
  	\bottomrule
        \toprule
            &Bank of Beijing& Ping An Insurance& NewChinaLife Insurance& Bank of Communications \\
        \midrule
        $B$ &166& 165& 141& 102 \\
		$B_{N}$  &0.1536& 0.1526& 0.1304& 0.0944 \\
  	\bottomrule
	\end{tabular}
	\label{tab:betweenness_zh}
\end{table}
\begin{table}[H]
	\centering
	\caption{The top 8 stocks by vertex betweenness centrality in the U.S. network.}
	\tabcolsep=0.75cm
	\begin{tabular}{ccccccc}
		\toprule
		  & Prudential Financial& Metlife& Loews& Wells Fargo  \\
		\midrule
        $B$ & 84& 67& 67& 39 \\
		$B_{N}$  & 0.1333& 0.1063& 0.1063& 0.0619 \\
  	\bottomrule
        \toprule
            &Aflac& Hartford Insurance& Globe Life& AIG \\
        \midrule
        $B$ &35& 21& 11& 5 \\
		$B_{N}$  &0.0556& 0.0333& 0.0175& 0.0079 \\
  	\bottomrule
	\end{tabular}
	\label{tab:betweenness_en}
\end{table}

Further comparing Tables~\ref{tab:betweenness_zh} and \ref{tab:betweenness_en}, we observe structural differences between the two financial systems. The top 8 central financial institutions in the Chinese market have significantly higher betweenness than those in the U.S. market, highlighting the more layered and hierarchical structure of the Chinese system. 
In the U.S. extremal dependence network, Prudential Financial, MetLife, and Loews show high betweenness centralities. This may be attributed to their diversified business models, spanning both insurance and other financial services. However, AIG's betweenness is only 0.0079, suggesting that, despite its size, its position in the network becomes less critical for the flow of risks. This may reflect its post-crisis restructuring and reduced involvement in high-risk financial products.

Hence, given the importance of betweenness centrality, we will use it as a key criterion in forming portfolios to mitigate systemic risks in the next section.



\subsection{Stock portfolio based on extremal dependence networks and maximum independent set}\label{subsec:portfolio}

Given the structure revealed by the extremal dependence network, regulators and investors may want to reduce contagion by isolating institutions with no direct connections, thus improving diversification and increasing systemic stability.
We achieve this goal by finding the \emph{maximum independent set} of the extremal dependence network. Since \emph{delta conditional value-at-risk} quantifies the systemic risk associated with the distress of a financial institution \cite{CoVaR}, we then use this measure to examine the effectiveness of the proposed MIS-based strategy.

In graph theory, for an undirected graph $G=(V,E)$, if $V^*\subseteq V$ and any two vertices in $V^*$ are not connected, then $V^*$ forms an independent set in graph $G$. If $V^*$ is not contained in any other independent set, it is called a maximal independent set. If the size of $V^*$ is the largest among all maximal independent sets, it is referred to as the maximum independent set. The maximum independent set problem (MISP) is a classic combinatorial optimization problem in graph theory. 

Finding an exact solution to MISP for a given graph has been shown to be NP-hard (cf. \cite{karp2010reducibility}). Therefore, as the size of the graph increases, the time complexity of solving MISP also increases, rendering exact solutions impractical. As a result, many researchers have developed heuristic-based approximate algorithms to solve the MISP (cf. \cite{palubeckis2008recursive, peng2015performance, wu2012multi}). Although these algorithms cannot guarantee optimal solutions, they can search much faster and guarantee cost-effectiveness when solving large-scale MISP problems. Currently, some of the most widely used heuristic algorithms include the greedy algorithm \cite{palubeckis2008recursive}, local search \cite{peng2015performance}, and Tabu search \cite{wu2012multi}. 
Here we use the greedy algorithm in \cite{palubeckis2008recursive} to find solutions to the MISP. This algorithm finds the solution set by gradually expanding the vertex set until all possibilities are exhausted to obtain a feasible solution. 

However, it is worthwhile noting that the result of the MIS may not be unique, since there may exist multiple sets of vertices that are not directly linked to each other and are equally maximal in size. Considering the discussion on centrality in Section~\ref{subseck:btw}, we prioritize stocks with lower betweenness centrality scores to further minimize systemic risk.

\begin{figure}[H]
	\centering
	\subcaptionbox{Chinese A-shares}{
		\begin{minipage}[h]{.45\linewidth}
			\centering
			\includegraphics[scale=0.3]{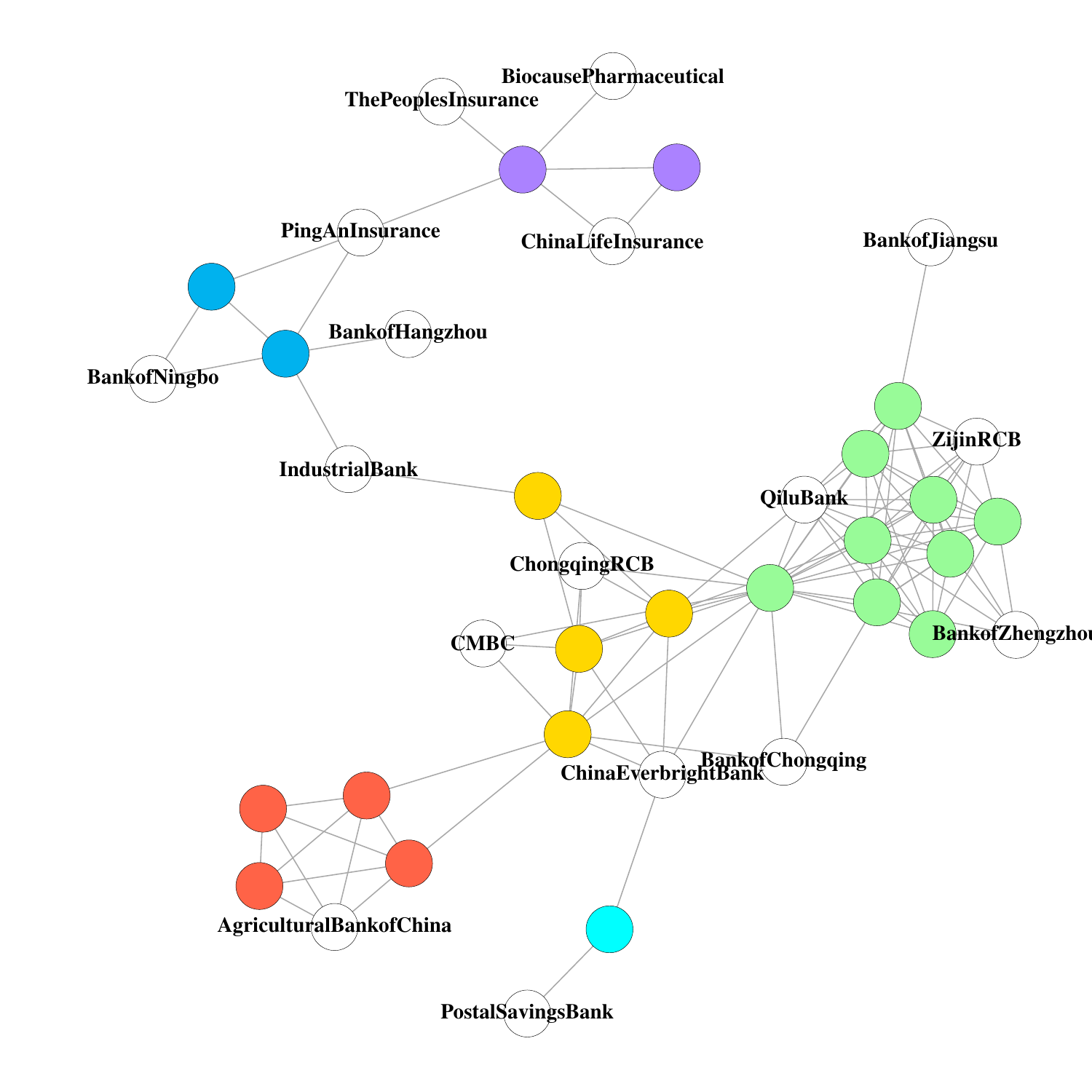}
		\end{minipage}
	}
	\subcaptionbox{U.S. S\&P 500}{
		\begin{minipage}[h]{.45\linewidth}
			\centering
			\includegraphics[scale=0.3]{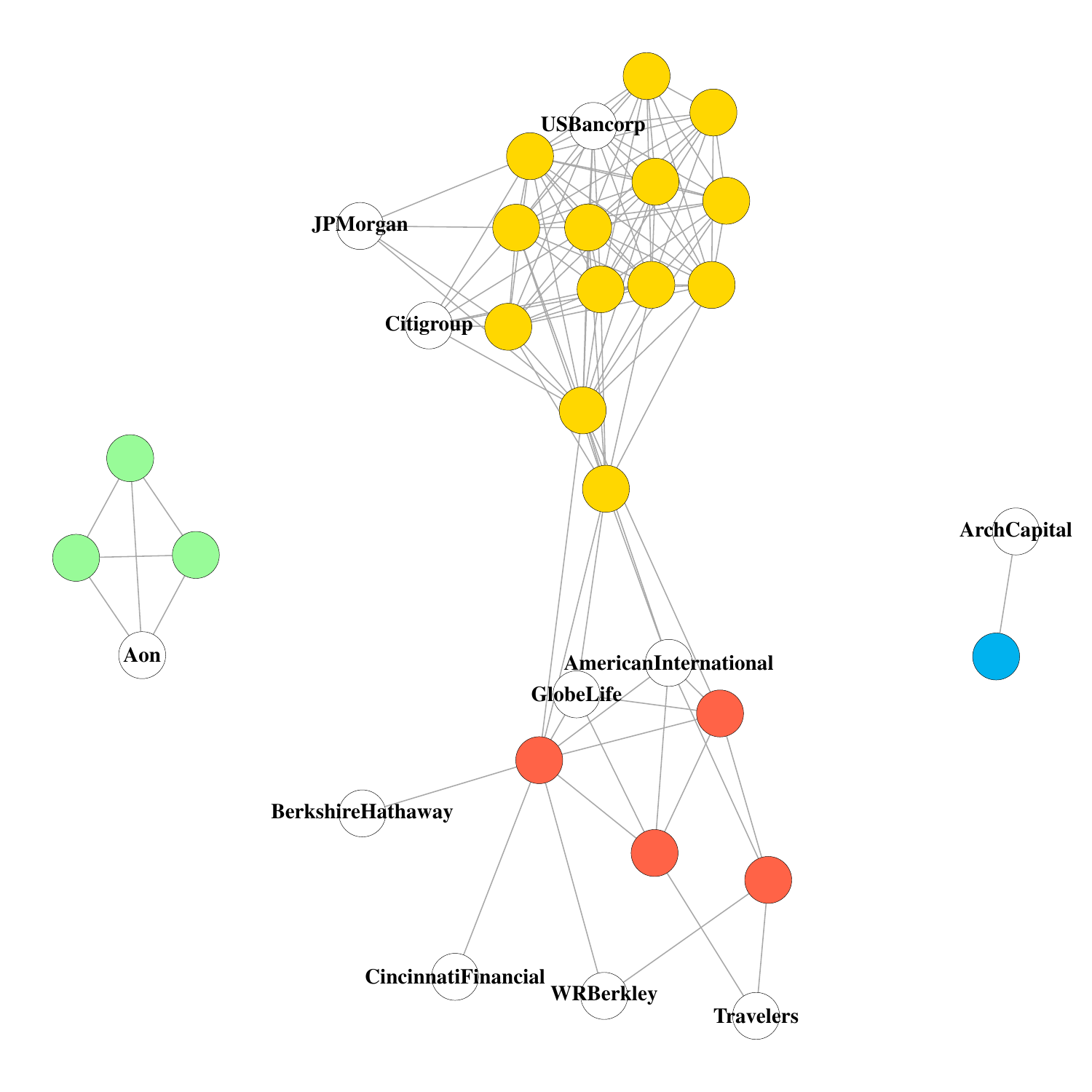}
		\end{minipage}
	}
	\caption{The MIS of each network graph, and they are denoted by white nodes.}
	\label{fig:MIS}
\end{figure}

We also highlight that due to the choice of threshold, isolated vertices in our extremal dependence networks do not necessarily indicate asymptotic independence from the rest of the institutions. When the threshold is low and the network becomes excessively redundant, unimportant relationships between these vertices may emerge, obscuring the main structure. In line with the discussion in Section~\ref{sec3.2}, we avoid overly complex networks and therefore exclude isolated vertices in our MIS strategy to prevent introducing misleading effects into the analysis.
The final portfolio of financial institutions chosen by the MIS strategy is given in \reffig{fig:MIS}, labeled with white nodes.

Note that in the literature, value-at-risk (VaR) is defined as (see for instance \cite{VaR})
\begin{equation*}
    \mathbb{P} (\Delta P < -{\text{VaR}}) = \alpha ,
\end{equation*}
where $\Delta P = P(t+\Delta t) - P(t)$ represents the loss over the holding period of length $\Delta t$.
To examine the effectiveness of risk mitigation for our MIS-based strategy, we evaluate its systemic risk exposure using $\Delta $CoVaR, which we now explain.
Let $X^i$ be the random variable for which $VaR^i$ is defined, i.e. 
\[
\mathbb{P}(X^i \le \text{VaR}^i_q) = q,
\]
and $C(X^i)$ denote some event related to institution $i$. 
Following \cite[Definition 1]{CoVaR}, $\text{CoVaR}^{i|C(X^j)}_q$ is defined as the $\text{VaR}$ of stock $i$ conditional on the event $C(X^j)$. That is, $\text{CoVaR}^{i|C(X^j)}_q$ corresponds to the $100q\%$-quantile of the conditional probability distribution such that
\begin{equation*}
    \mathbb{P} \left ( X^i|C(X^j) \leq \text{CoVaR}^{i|C(X^j)}_q \right ) = q.
\end{equation*}
In \cite{CoVaR}, the authors attribute the systemic risk contribution of stock $j$ to stock $i$ by
\begin{equation}\label{eq:deltaCoVaR}
    \Delta \text{CoVaR}^{i|j}_q = \text{CoVaR}^{i|X^j = \text{VaR}^j}_q - \text{CoVaR}^{i|X^j = \text{VaR}^j_{50}}_q.
\end{equation}
In particular, by \eqref{eq:deltaCoVaR}, if asymptotic independence holds between $X^i$ and $X^j$, then both $\Delta \text{CoVaR}^{i|j}_q$ and $\Delta \text{CoVaR}^{j|i}_q$ are equal to 0.

\begin{figure}[h]
	\centering
	\subcaptionbox{Chinese A-shares}{
		\begin{minipage}[h]{.48\linewidth}
			\centering
			\includegraphics[scale=0.3]{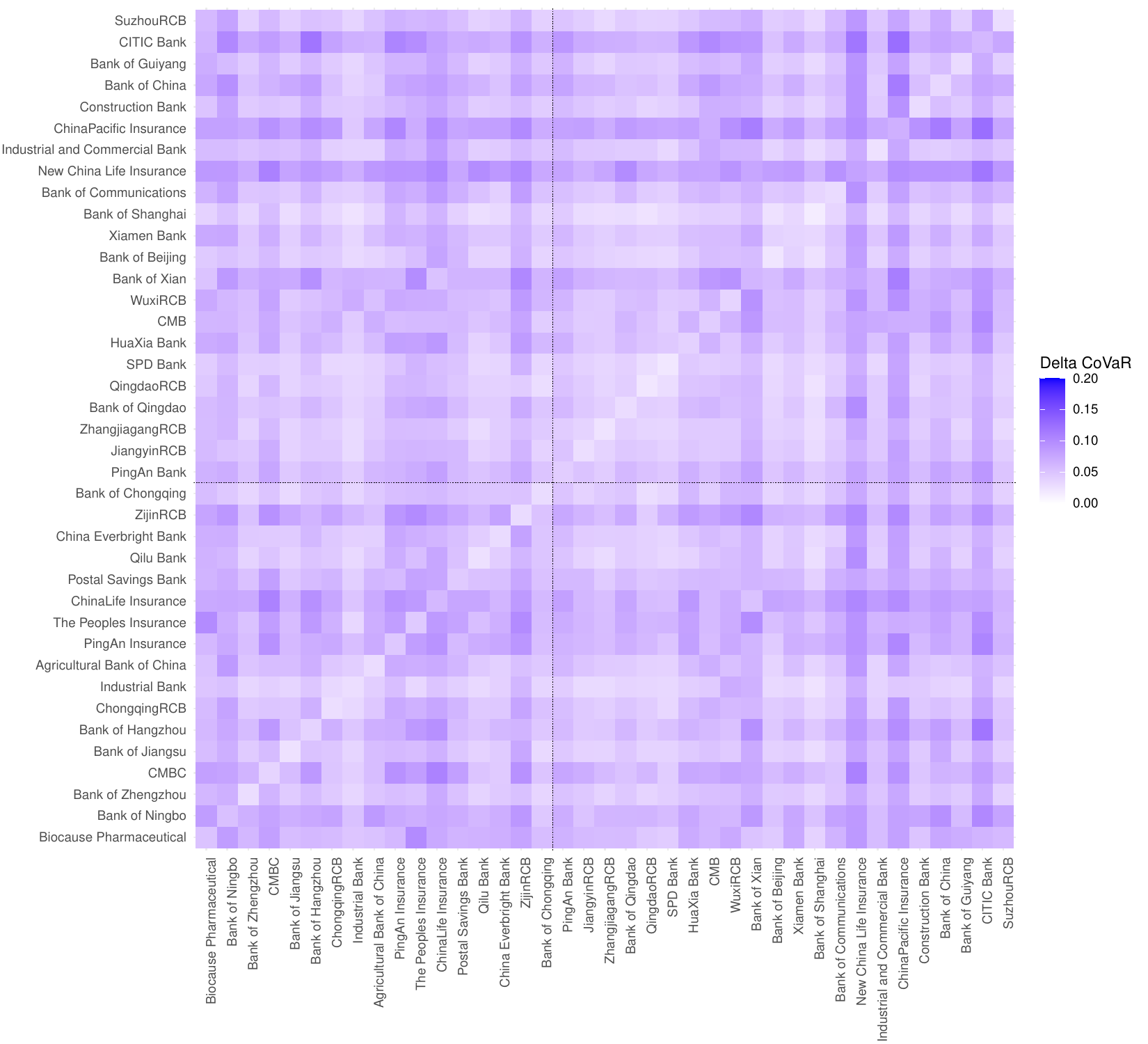}
		\end{minipage}
	}
	\subcaptionbox{U.S. S\&P 500}{
		\begin{minipage}[h]{.48\linewidth}
			\centering
			\includegraphics[scale=0.3]{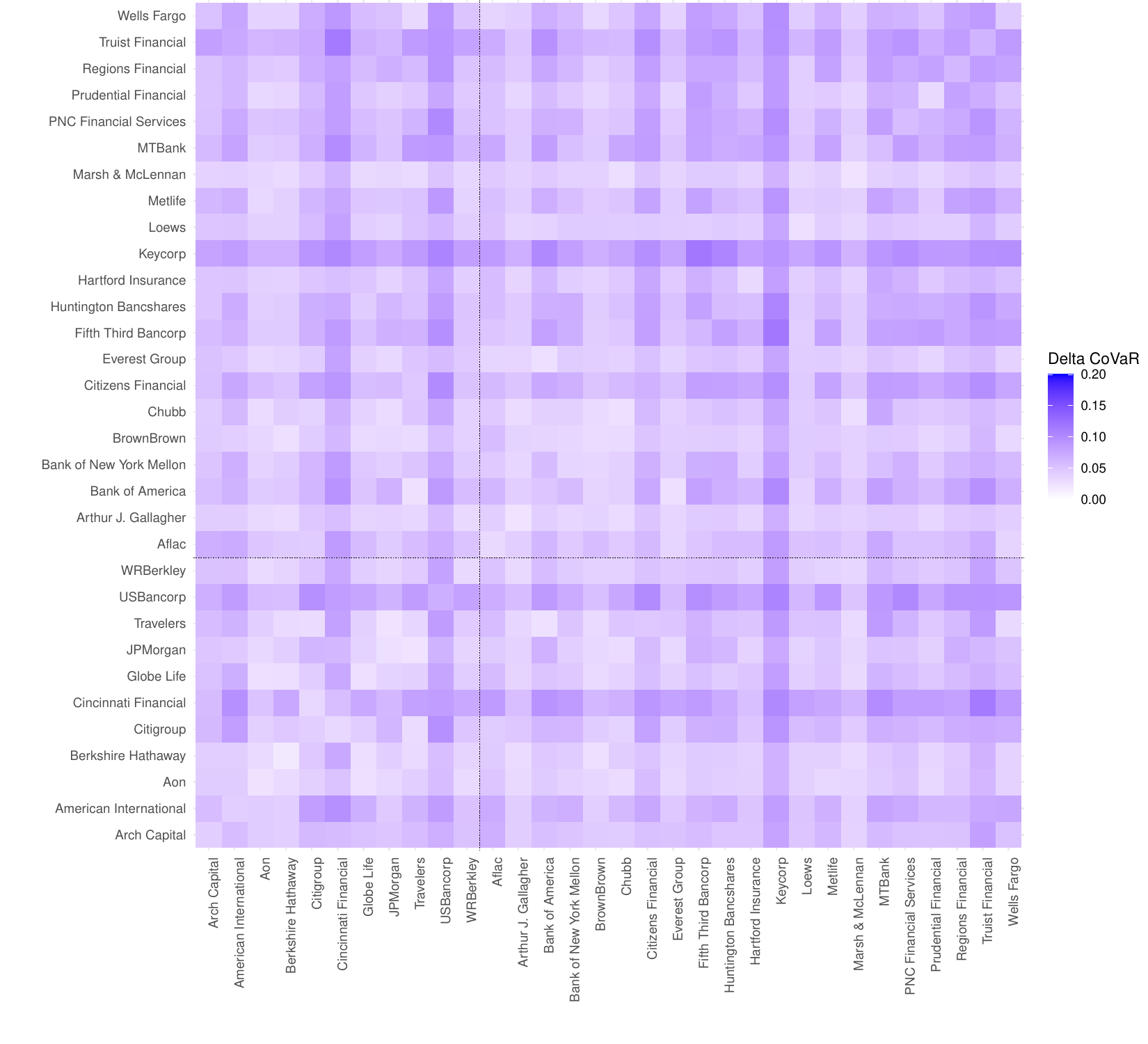}
		\end{minipage}
	}
	\caption{Heatmaps of $(\Delta \text{CoVaR}^{i|j} + \Delta \text{CoVaR}^{j|i})/2$ for each pair of $i$ and $j$. The financial institutions are ordered such that those chosen by the MIS method come first. Hence, the lower left blocks depict the systemic risks of the chosen portfolios in the two systems.}
	\label{fig:delta_CoVaR}
\end{figure}

To proceed, we set $q=0.99$ and calculate the $\Delta $CoVaR for each pair of stocks. Then we generate heatmaps based on the average effects, $(\Delta \text{CoVaR}^{i|j} + \Delta \text{CoVaR}^{j|i})/2$, and results are given in \reffig{fig:delta_CoVaR}. 
We also order the financial institutions so that those chosen by the MIS method are listed first.
Note that both the Chinese and U.S. maps are plotted on the same color scale to make direct comparisons between the two financial systems.

In both systems, most of the institutions chosen by the MIS method (i.e. the lower left blocks in the heatmaps) tend to have lower systemic risk. However, the two heatmaps also help identify a few influential institutions in the system. For example, in \reffig{fig:delta_CoVaR}(a), institutions such as New China Life Insurance and China Pacific Insurance exhibit higher average $\Delta $CoVaR values with other stocks, highlighting potential risk centers. The pivotal role of New China Life Insurance also coincides with the conclusion drawn from the betweenness centrality.

Meanwhile, \reffig{fig:delta_CoVaR}(b) shows that KeyCorp and USBancorp are two potentially risky spots in the U.S. financial system. Notably, Fitch Ratings downgraded KeyCorp's default ratings in October 2023, reflecting concerns about KeyCorp's credit quality. Also, we see that institutions with high betweenness centrality scores do not reveal concerning exposure to systemic risk, showing that the U.S. regulations on financial institutions are now effective.



\section{Empirical study and results}\label{sec4}
In this section, we propose a portfolio strategy to minimize the risk of extremal loss, where one common systemic risk measure, expected shortfall (ES) is used. We also compare our portfolios with the main market indices in both China and the U.S., offering investment advice tailored to different levels of risk tolerance.

\subsection{Optimal portfolio with minimum risk}
Expected shortfall (ES) is a \emph{coherent} risk measure which satisfies the four important axioms: translation invariance, subadditivity, positive homogeneity, and monotonicity \cite{ES}. Here we use ES to measure the risk of the portfolio.
According to \cite{ES}, ES is defined as
\begin{equation}
    \label{eq:def_ES}
	{\text{ES}} = \mathbb{E}[L|L>{\text{VaR}}],
\end{equation}
which quantifies the expected loss in adverse scenarios.

We assume the holding period $\Delta t$ is one day and calculate the ES of each stock at a $95\%$ confidence level. The objective function is to minimize the overall risk (ES) of the portfolio. Constraints are imposed to ensure that the sum of weights equals 1, with each weight coefficient ranging from 0 to 0.1. 
For the Chinese market, given that the three‐month deposit rate in 2024 set by the Chinese Central Bank is $1.15\%$, we require the expected return of the portfolio to be at least $1.15\%$. This problem is then formulated as a linear programming problem.
 Let $|\text{MIS}_\text{China}|$ and $|\text{MIS}_\text{US}|$ denote the cardinality of the two identified maximum independence sets, and suppose $\text{ES}_i$, $c_i$, and $R_i$ refer to the ES, weight, and return of stock $i$, respectively. Then for the Chinese system, we need to solve the following optimization problem:
\begin{equation}\label{eq:portfolio}
    \begin{split}
        &\min \ \ \sum\limits_{i=1}^{|\text{MIS}_\text{China}|} c_i \text{ES}_i \\
        &s.t. \left\{\begin{array}{lc}
            \sum\limits_{i=1}^{|\text{MIS}_\text{China}|} c_i=1 \\
             0\leq c_i \leq 0.1, \quad i=1,\dots, |\text{MIS}_\text{China}| \\
             \sum\limits_{i=1}^{|\text{MIS}_\text{China}|} c_iR_i \geq 1.15\%.
        \end{array}\right.
    \end{split}
\end{equation}
A similar formulation is adopted for the U.S. market, with the return lower bound adjusted to $0.2\%$:
\begin{equation}\label{eq:portfolio_en}
    \begin{split}
        &\min \ \ \sum\limits_{i=1}^{|\text{MIS}_\text{US}|} c_i \text{ES}_i \\
        &s.t. \left\{\begin{array}{lc}
             \sum\limits_{i=1}^{|\text{MIS}_\text{US}|} c_i=1 \\
             0\leq c_i \leq 0.1,\quad i=1,\ldots, |\text{MIS}_\text{US}| \\
             \sum\limits_{i=1}^{|\text{MIS}_\text{US}|} c_iR_i \geq 0.2\%,
        \end{array}\right.
    \end{split}
\end{equation}

For each vertex in the MIS, we compute their expected shortfalls, then solve the optimization problems in \eqref{eq:portfolio} and \eqref{eq:portfolio_en} to obtain the optimal portfolio weights by using the \verb6linprog6 function in MATLAB. The results are summarized in Tables~\ref{tab:weight_zh} and \ref{tab:weight_en}.

\begin{table}[h]
	\centering
	\caption{Optimal portfolio with the minimum risk for the maximum independent set in Chinese A-shares; the obtained objective function value is 2.17\% (ES), and the total return rate is 1.15\%.}
	\tabcolsep=0.3cm
	\begin{tabular}{ccccccc}
		\toprule
		  & Biocause      & Bank of& Bank of&\multirow{2}{*}{CMBC} &Bank of&Bank of \\
            & Pharmaceutical& Ningbo& Zhengzhou&                     &Jiangsu&Hangzhou  \\
		\midrule
		ES (\%) & 3.1850& 3.7529& 1.9250& 2.3316& 1.7491& 2.2230\\
        Weight (\%) & 0& 0& 10.00 & 10.00& 10.00& 10.00 \\
  	\bottomrule
        \toprule
		  & \multirow{2}{*}{ChongqingRCB}& Industrial& Agricultural&PingAn &The Peoples&ChinaLife \\
            &                            & Bank       & Bank of China&Insurance&Insurance&Insurance  \\
		\midrule
		ES (\%) & 2.1066& 1.9503& 2.5589& 3.0791& 2.7382& 3.8722\\
        Weight (\%) & 3.23& 10.00& 0 & 0& 6.77& 0 \\
  	\bottomrule
        \toprule
		  & Postal Savings& \multirow{2}{*}{Qilu Bank}& China Everbright&\multirow{2}{*}{ZijinRCB} &Bank of& \\
            & Bank          &                           & Bank           &                     &Chongqing&  \\
		\midrule
		ES (\%) & 3.4818& 1.8579& 2.4800& 2.3637& 2.3155& \\
        Weight (\%) & 0& 10.00& 10.00 & 10.00& 10.00&  \\
  	\bottomrule
	\end{tabular}
	\label{tab:weight_zh}
\end{table}
\begin{table}[h]
	\centering
	\caption{Optimal portfolio with the minimum risk for the maximum independent set in U.S. S\&P 500; the obtained objective function value is 3.14\% (ES), and the total return rate is 0.28\%.}
	\tabcolsep=0.28cm
	\begin{tabular}{ccccccc}
		\toprule
		  & \multirow{2}{*}{Arch Capital}& American & \multirow{2}{*}{Aon}& \multirow{2}{*}{Berkshire Hathaway}& \multirow{2}{*}{Citigroup}&Cincinnati  \\
          & &International& & & &Financial \\
		\midrule
		ES (\%) & 3.8608& 4.0205 & 3.2939 & 1.8852& 3.5773& 2.9820\\
        Weight (\%) & 10.00& 10.00& 10.00& 10.00& 10.00& 10.00\\
  	\bottomrule
        \toprule
		  & Globe Life& JPMorgan& Travelers& USBancorp& WRBerkley& \\
		\midrule
		ES (\%) & 2.8204&  2.9781&  2.8766& 5.8912& 3.1252& \\
        Weight (\%) & 10.00& 10.00& 10.00& 0& 10.00& \\
  	\bottomrule
	\end{tabular}
	\label{tab:weight_en}
\end{table}

As revealed by Tables~\ref{tab:weight_zh} and \ref{tab:weight_en}, the ES values for the chosen stocks in the Chinese A-shares (ranging from 1.75\% to 3.75\%) are substantially lower than those in the U.S. S\&P 500 (varying from 1.89\% to 5.89\%), indicating a less risky profile for the Chinese market. 

Table~\ref{tab:weight_zh} also shows institutions such as the Agricultural Bank of China, PingAn Insurance, China Life Insurance, and Postal Savings Bank are assigned a weight of zero. These institutions are located in the darker zone of the lower-left block of \reffig{fig:delta_CoVaR}(a), which indicates higher systemic risk. 
In Table~\ref{tab:weight_en}, only the USBanCorp is assigned zero weight. As highlighted in \reffig{fig:delta_CoVaR}(b), USBanCorp is one of the two institutions with higher systemic risk. Therefore, the zero weight assignments in both Chinese and U.S. portfolios further confirm the effectiveness of our method in avoiding investments in higher-risk institutions.

Moreover, the entire Chinese portfolio exhibits a lower ES of 2.17\% with a less uniform weight distribution compared to the U.S. portfolio, which achieves a portfolio ES of 3.14\%, and stocks other than the USBanCorp all receive an equal weight of 10\%. 
 This suggests that while the Chinese market in general shows lower systemic risk, certain institutions still have high-risk exposures. The skewed allocation in the Chinese portfolio highlights the need for selective weighting, even in lower-risk markets, to avoid concentrated vulnerabilities. These findings emphasize the importance of a tailored approach to risk management, considering both overall market risk and individual institutions' exposures.

\subsection{Performance Comparison of MIS Portfolios and Main Markets}
\begin{figure}[h]
	\centering
	\includegraphics[width=\textwidth]{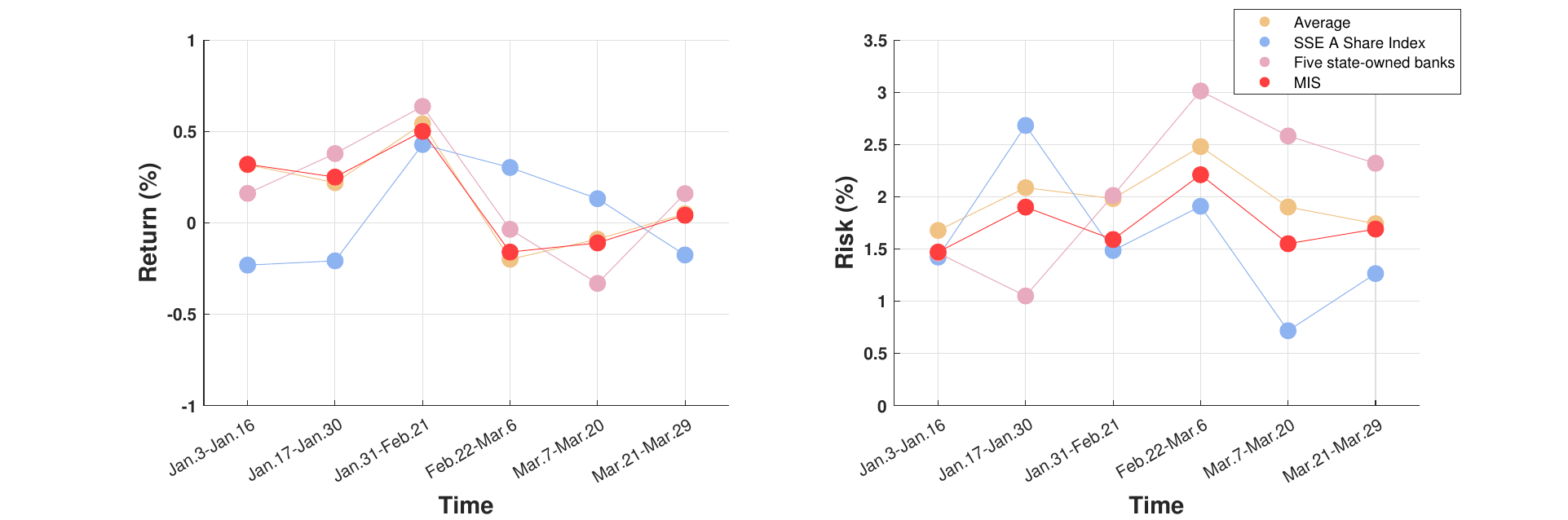}
	\caption{Performance comparison of the MIS portfolio, average return of bank and insurance stocks, SSE A-share Index, and five state-owned banks.}
	\label{fig:A-share}
\end{figure}

\begin{figure}[h]
	\centering
	\includegraphics[width=\textwidth]{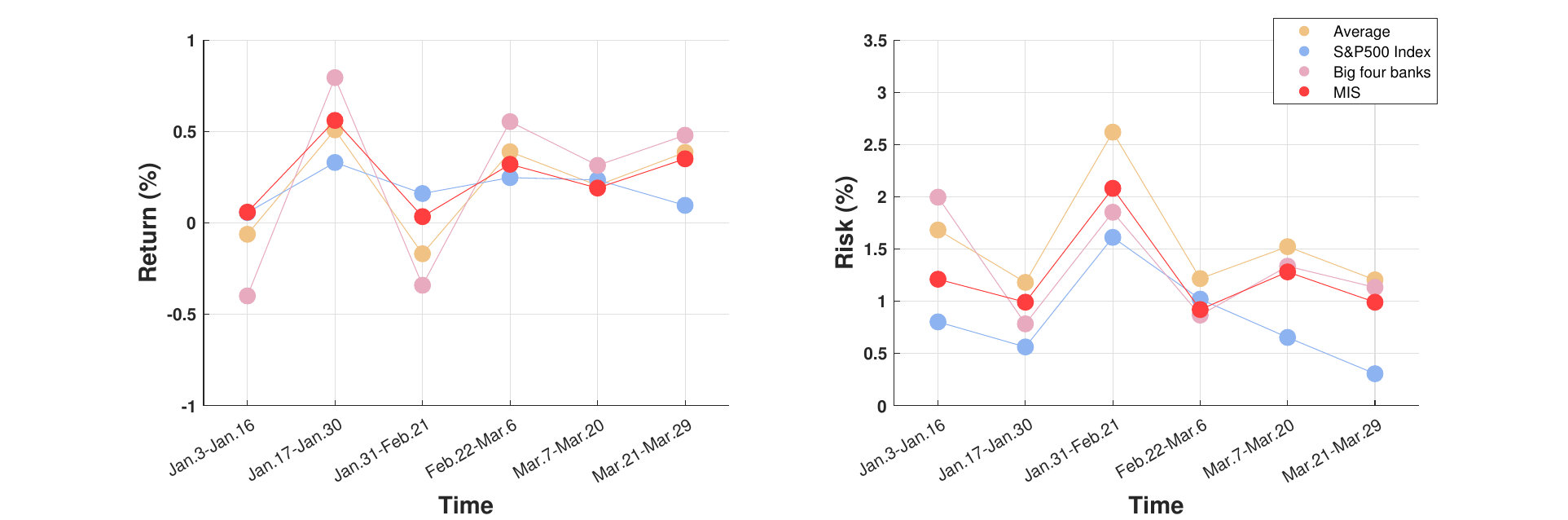}
	\caption{Performance comparison of the MIS portfolio, average return of bank and insurance stocks, S\&P 500 Index, and big four banks.}
	\label{fig:SP500}
\end{figure}

To further assess the performance of our stock portfolios in 2024, we obtain stock prices from January 3rd to March 29th, 2024, and segment them into six intervals of 10 trading days each. We compare the actual portfolio returns with the market portfolio (A-shares and S\&P 500), and results are presented in Figures~\ref{fig:A-share} and \ref{fig:SP500}. Specifically, we evaluate the maximum independent set (MIS) of bank and insurance stocks, the average return of all bank and insurance stocks, the corresponding market indices (SSE A-share Index and S\&P 500 Index), and the largest banks of each market.

In Figure~\ref{fig:A-share}, we observe that the SSE A-Share Index, representing the broad market, exhibits a volatile return pattern over the selected time interval, potentially reflecting economic uncertainty and cautious investor sentiment. Meanwhile, the MIS portfolio and the average return of all Chinese bank and insurance stocks generally follow comparable trends, but the MIS portfolio shows a more stable trend, particularly during January 2024 when the SSE A Share Index records negative returns.
In addition, the portfolio consisting of the five state-owned banks displays a return profile that closely aligns with the MIS portfolio but is more volatile, especially between March 7 and March 20, 2024.
Hence, the MIS portfolio gives a competitive return profile that mirrors the performance of the five state-owned banks. This alignment may reflect the potential influence of government intervention and the hierarchical structure within the Chinese financial system.
In terms of risk, the SSE A-Share Index shows higher fluctuations compared to the more stable risk profiles of the MIS portfolio. The peak in returns across most portfolios around the January 31 - February 21 interval suggests a potentially significant market event or shift during that time. The relatively stable risk of the MIS indicates a more conservative investment approach compared to the broader market index.

Moreover, we note from Figure~\ref{fig:SP500} that in the U.S. market, the S\&P 500 Index shows a stable return pattern with lower risk during the chosen period compared to the other three portfolios. The average returns of the big four banks, as well as those of all U.S. bank and insurance company stocks, are the most volatile between January and March 2024. This difference may be attributed to the banking sector is still facing ongoing challenges of the 2023 regional banking crisis, during which institutions like Silicon Valley Bank collapsed, whereas the S\&P 500 benefited from diversification and tech-driven stability. 
However, among the four portfolios, the MIS portfolio achieves a balance between high returns and low risk. Its returns outperform the market index most of the time, while maintaining a more stable risk profile than both the big four U.S. banks and the broad financial and insurance sector. This suggests that the MIS portfolio offers a more resilient investment option during periods of financial sector volatility.

The analysis highlights different risk-return dynamics in the Chinese and U.S. markets during the first quarter of 2024. The Chinese market index exhibits high volatility during the selected time period, while the MIS portfolio remains more stable, closely aligning with state-owned banks, possibly due to government influence. In the U.S., the S\&P 500 shows steady returns over, but financial sector stocks experience high volatility following the 2023 banking crisis. Across both markets, the MIS portfolio demonstrates its effectiveness in consistently balancing high returns with lower risk. Its stability during periods of market and sector uncertainties reveals its resilience as a strategic investment choice to mitigate volatility and preserve value.

 

\section{Conclusions}\label{sec5}
In this paper, we employ the classical extremal dependence measure to quantify the dependence between financial institution stocks and represent their relationships as a network. Using data from 2023, we analyze 48 Chinese A-shares and 37 U.S. S\&P 500 stocks to compare the systemic risk structures of the two markets. By investigating key network characteristics, we identify unique properties in the Chinese and U.S. financial systems, offering insights into how systemic risk propagates and can be mitigated. We also construct portfolios based on the maximum independent set (MIS), aiming to minimize expected shortfalls. Finally, we assess the real-world performance of these constructed portfolios in the first quarter of 2024, demonstrating the effectiveness of our approach in both markets.

For future work, one may consider incorporating dynamic network models to enhance the adaptability of our methods to evolving market conditions. Also, integrating our methodology with graph neural networks (GNNs) can enable more sophisticated risk modeling using deep learning techniques to detect complex dependence structures in volatile financial networks.

\section*{Acknowledgments}
T. Wang gratefully acknowledges the National Natural Science Foundation of China Grant 12301660 and the Science and Technology Commission of Shanghai Municipality Grant 23JC1400700.

Both authors also thank Shanghai Institute for Mathematics and Interdisciplinary Sciences (SIMIS) for their financial support. This research was partly funded by SIMIS under grant number SIMIS-ID-2024-WE. We are grateful for the resources and facilities provided by SIMIS, which were essential for the completion of this work.


\bibliographystyle{plainnat}
\bibliography{ref.bib}

\end{document}